# Predicting Mesoscopic Larmor Frequency Shifts in White Matter with Diffusion MRI - A Monte-Carlo Study in axonal phantoms

*Anders Dyhr Sandgaard[1], Sune Nørhøj Jespersen[1,2]*

[1]Center of Functionally Integrative Neuroscience, Department of Clinical Medicine, Aarhus University, Denmark

[2]Department of Physics and Astronomy, Aarhus University, Denmark

**Corresponding Author**: Anders Dyhr Sandgaard.

**Postal address**: CFIN, Department of Clinical Medicine, Universitetsbyen 3, Building 1710, 8000 Aarhus C, Denmark.

**Mail**: anders@cfin.au.dk

## Abbreviations:

**fODF:** Fiber orientation Distribution Function. **dMRI**: Diffusion MRI. **SM**: Standard Model of Diffusion in White Matter. **MGE**: Multi-Gradient echo signal. **QSM**: Quantitative Susceptibility Mapping. **STI**: Susceptibility Tensor Imaging. **μQSM**: Microstructure-informed QSM. **MC**: Monte-Carlo. **WM**: White Matter. **GM**: Gray Matter. **TBI**: Traumatic Brain Injury. **ASE**: Asymmetric Spin Echo. **PGSE**: Pulsed-Gradient Spin Echo. **EM**: Electron Microscopy. **CC**: Corpus Callosum. **CG**: Cingulum Bundle. **Contra**: Contralateral. **Ipsi**: Ipsilateral. **COM**: Center of mass. **SE**: Spin Echo. **BIC**: Bayesian Information Criterion. **NRMSE**: Normalized Root-Mean-Square-Error. **SD**: Standard Deviation.

## Symbols:

$\hat{\boldsymbol{n}}$: Cylinder direction vector

$\mathcal{P}^{\mathrm{D}}(\hat{\boldsymbol{n}})$: Fiber Orientation Distribution Function (fODF) estimated with diffusion MRI

$\mathcal{P}^{\mathrm{EM}}(\hat{\boldsymbol{n}})$: Fiber Orientation Distribution Function (fODF) estimated from Electron microscopy

$\mathcal{P}^{\Omega}(\hat{\boldsymbol{n}})$: Fiber Orientation Distribution Function (fODF) estimated from the Larmor frequency shift

**T**: Mean orientation tensor of cylinder directions $\hat{\boldsymbol{n}}$. Relates to second moment of $\mathcal{P}(\hat{\boldsymbol{n}})$

$p_{2m}^{\mathrm{EM}}$: Expansion coefficients of $\mathcal{P}^{\mathrm{EM}}(\hat{\boldsymbol{n}})$ in spherical harmonics $Y_l^m(\hat{\boldsymbol{n}})$

$p_{2m}^{\mathrm{D}}$: Expansion coefficients of $\mathcal{P}^{\mathrm{D}}(\hat{\boldsymbol{n}})$ in spherical harmonics $Y_l^m(\hat{\boldsymbol{n}})$

$p_{2m}^{\Omega}$: Expansion coefficients of $\mathcal{P}^{\Omega}(\hat{\boldsymbol{n}})$ in spherical harmonics $Y_l^m(\hat{\boldsymbol{n}})$

$\chi$: Magnetic Susceptibility

$\delta\chi$: Demeaned magnetic susceptibility

$\mathbf{B}_0 = \hat{\mathbf{B}}B_0$: Magnetic field vector with direction $\hat{\mathbf{B}}$ and magnitude $B_0$

$\gamma$: Gyromagnetic ratio of water.

$\boldsymbol{\Upsilon}(\boldsymbol{r})$: Dipole field tensor

$\Delta\mathbf{B}(\boldsymbol{r})$: Induced magnetic field of tissue

$\Omega(\boldsymbol{r})$: Local Larmor frequency shift induced by $\Delta\mathbf{B}(\boldsymbol{r})$

$\overline{\Omega}$: Spatially averaged Larmor frequency shift described by $\Delta\overline{\mathbf{B}}$ across all water

$\overline{\Omega}_i$: Spatially averaged Larmor frequency shift described by $\Delta\overline{\mathbf{B}}$ across water in the $I$'th compartment

$\overline{\Omega}_{\mathrm{MGE}}$: Measured Larmor frequency shift of multi gradient-echo signal

$\overline{\Omega}_{\mathrm{ASE}}$: Measured Larmor frequency shift of asymmetric spin echo.

$\overline{\Omega}^{\mathrm{Meso}}$: Mesoscopic contribution to the measured Larmor frequency from the local magnetic microstructure

$\overline{\Omega}^{\mathrm{Macro}}$: Macroscopic contribution to the measured Larmor frequency by tissue from neighboring voxels

$D_0$: Free diffusivity of water

$D_a$: Intra-axonal diffusivity of water

$W_a^{\parallel}$: Intra-axonal axial kurtosis

$\Delta$: Diffusion time

$v_a(\boldsymbol{r})$: Spatial indicator function of intra-axonal water

$\zeta_a$: Volume fraction of intra-axonal water

$v_e(\boldsymbol{r})$: Spatial indicator function of myelin sheath

$\zeta_e$: Volume fraction of myelin sheath

$\mathbf{A}(\boldsymbol{r})$: Magnetic field tensor from induced tissue magnetic field, so $\Omega(\boldsymbol{r}) = \gamma B_0 \hat{\mathbf{B}}^T \mathbf{A}(\boldsymbol{r}) \hat{\mathbf{B}}$

$\delta_l$: Step length in Monte-Carlo simulation

$\delta_t$: Time step for every $\delta_l$

$S_{\text{MGE}}$: Monte-Carlo simulated multi gradient-echo signal

$S_{\text{ASE}}$: Monte-Carlo simulated asymmetric spin-echo signal

$S_{\text{PGSE}}$: Monte-Carlo simulated pulsed-gradient spin-echo signal

$S_{SM}$: Signal of Standard Model of Diffusion

$\boldsymbol{\varphi}$: Induced signal phase tensor caused by $\mathbf{A}$

$T_E$: Echo time

$\Delta T_E$: Delayed time after $T_E$ of signal read-out

$\alpha_{T_E}$: 180-degree RF pulse indicator function

$\boldsymbol{r}_p$: Spatial position vector of particle $p$

$\Delta \mathbf{l}_p$: Displacement vector of particle $p$ between the two diffusion gradient pulses

$\boldsymbol{r}_{p0}$: Initial position vector of particle $p$

$\delta \boldsymbol{r}_p(t)$: Position vector of particle $p$ after time $i$

$\mathbf{g}$: Diffusion gradient with direction $\hat{\mathbf{g}}$ and b-value $b$

$\delta$: Duration diffusion gradient is turned on in each $\mathbf{g}$ pulse

$\mathcal{K}$: Signal Kernel for a bundle of parallel sticks

NRMSE: Normalized Root-Mean-Square-Error between estimated and simulated Larmor frequency

$\beta_D$: Estimated error ratio between fitted susceptibility and ground truth using $p_{2m}^D$

$\beta_{\text{EM}}$: Estimated error ratio between fitted susceptibility and ground truth using $p_{2m}^{\text{EM}}$

# 1| Abstract

Magnetic susceptibility MRI offers potential insights into the chemical composition and microstructural organization of tissue. However, estimating magnetic susceptibility in white matter is challenging due to anisotropic sub-voxel Larmor frequency shifts caused by axonal microstructure relative to the B0 field orientation. Recent biophysical models have analytically described how axonal microstructure influences the Larmor frequency shifts, relating these shifts to a mesoscopically averaged magnetic field that depends on the axons' fiber Orientation Distribution Function (fODF), typically estimated using diffusion MRI. This study aims to validate the use of MRI to estimate mesoscopic magnetic fields and determine whether diffusion MRI can faithfully estimate the orientation dependence of the Larmor frequency shift in realistic axonal microstructure. To achieve this, we developed a framework for performing Monte-Carlo simulations of MRI signals in mesoscopically sized white matter axon substrates segmented with electron microscopy. Our simulations demonstrated that with careful experimental design, it is feasible to estimate mesoscopic magnetic fields. Additionally, the fODF estimated by the Standard Model of diffusion in white matter could predict the orientation dependence of the mesoscopic Larmor frequency shift. We also found that incorporating the intra-axonal axial kurtosis into the Standard Model could explain a significant amount of signal variance, thereby improving the estimation of the Larmor frequency shift. This factor should not be neglected when fitting the Standard Model.

# 2| Introduction

MRI is a powerful imaging technique sensitive to detecting water diffusion and microscopic magnetic field variations induced by the magnetized tissue[1,2]. This capability allows examination of tissue at scales significantly finer than the image resolution, enabling observation of voxel-averaged magnetic and microstructural tissue changes. Consequently, MRI is well-suited for detecting minor alterations in brain tissue, which are crucial for studying neurodegeneration. To achieve both sensitive and specific microstructural characterization, biophysical tissue models have been developed to identify subtle tissue changes.

In diffusion MRI (dMRI), the Standard Model of Diffusion in White Matter[2] (SM), which encompasses a range of previously utilized diffusion models, has become a popular biophysical model in brain tissue research[3–7]. Here, the intra-axonal and extra-axonal signal is modelled as Gaussian and non-exchanging compartments for long diffusion times. For example, the intra-axonal signal can be described as a Gaussian signal in a 1D stick compartment, with non-zero diffusivity only along its axis. The total signal from many non-exchanging axon bundles is then modelled as a weighted sum across the axonal directions, governed by a fiber orientation distribution function (fODF) $\mathcal{P}^D(\hat{n})$. Here, the "D" indicates it is estimated by diffusion MRI. The diffusivities reflect a coarse-grained version of the axonal geometry, while $\mathcal{P}^D(\hat{n})$ models the orientation of axonal bundles on the mesoscopic scale.

Biophysical models have also been developed to describe how the tissue magnetic susceptibility χ, describing the magnetization response in the presence of an external $\mathbf{B}_0$ field, shifts the measured Larmor frequency shift $\overline{\Omega}_{\text{MGE}}$ of the multi-gradient-echo (MGE) signal[8–14]. On the microscopic scale, susceptibility depends on the chemical composition and molecular structure of cells, providing biomarkers for iron[15–21], lipid, protein content[22–30], etc.

The magnetized tissue generates microscopically varying magnetic fields that can shift the Larmor frequency by several hertz in clinical scanners[14]. This field depends not only on susceptibility but also on tissue microstructure – encapsulated by the common term *magnetic microstructure*. While diffusion typically probes microstructure on the mesoscopic scale[31] (10-100 μm), the induced magnetic field is a non-local phenomenon, meaning that the entire tissue sample and its surroundings cause shifts in the spin's Larmor frequency at every microscopic position in the sample. Even after the massive averaging across the voxel, the measured Larmor frequency shift still contains a mesoscopic contribution $\bar{\Omega}^{\text{Meso}}$ that depends exclusively on local magnetic microstructure at distances below the imaging resolution. The mesoscopic contribution $\bar{\Omega}^{\text{Meso}}$ can be as significant as the frequency shift induced by tissue from neighboring voxels denoted $\bar{\Omega}^{\text{Macro}}$. Unfortunately, current susceptibility estimation techniques like Quantitative Susceptibility Mapping[32,33] (QSM) or Susceptibility Tensor Imaging[26] (STI) disregard any sub-voxel frequency shift $\bar{\Omega}^{\text{Meso}}$ and assume that the measured signal's Larmor frequency shift can be described by the induced shift $\bar{\Omega}^{\text{Macro}}$ from neighboring voxels. This approximation may appear unavoidable because incorporating microscopic shifts depends on explicit microstructure, leading to more parameters than what is feasible to estimate[34] - even after sampling at multiple $\mathbf{B}_0$ directions. However, if the additional microstructural degrees of freedom, that arise by modelling the sub-voxel frequency shifts, could be known prior to estimating susceptibility, the number of unknowns would be drastically reduced, leaving only a few rotation-invariant susceptibility-related parameters to be estimated[10,12,14,34].

Recently, an analytical framework was derived that accounts for both $\bar{\Omega}^{\text{Meso}}$ and $\bar{\Omega}^{\text{Macro}}$ when estimating magnetic susceptibility from MRI[9,12]. It was shown how $\bar{\Omega}^{\text{Meso}}$ depends on the structural correlation function of the microstructure when susceptibility is uniform. This framework was later used to develop microstructure informed QSM (μQSM) (previously known as QSM+), which incorporates mesoscopic frequency shifts from multi-layer cylinders with scalar susceptibility and arbitrary orientation dispersion into QSM. Both QSM, μQSM and STI model the measured Larmor frequency shift in MRI via the voxel-averaged magnetic field described by the first signal cumulant. In μQSM, it was derived how the orientation dependence of the mesoscopic Larmor frequency shift $\bar{\Omega}^{\text{Meso}}$ from dispersed cylinders is captured by a scatter matrix[35], $\mathbf{T} = \langle \hat{\mathbf{n}}\hat{\mathbf{n}}^{\mathbf{T}} \rangle$, representing the mean orientation tensor of the cylinder orientations $\hat{\mathbf{n}}$. $\mathbf{T}$ relates to the second moment of the fODF. Utilizing that $\mathbf{T}$ may also be estimated from the fODF with dMRI, it was demonstrated that accounting for $\bar{\Omega}^{\text{Meso}}$ substantially changed susceptibility values in WM in an ex vivo mouse brain.

For an ideal medium of cylinders, the anisotropy of the two different contrast mechanisms (diffusion and magnetic susceptibility) should depend on the same scatter matrix $\mathbf{T}$. However, real tissue deviates from this ideal picture, and it remains to be investigated how these two "orthogonal" modalities are differently affected by realistic axonal features such as beading, tortuosity, and undulations[36]. While the two modalities may be influenced by the same structural correlation function, their parameters might be biased differently due to model simplifications. It is therefore timely to investigate if $\mathbf{T}$ from dMRI is similar to $\mathbf{T}$ as seen through susceptibility induced Larmor frequency shifts.

One approach to comparing structural information across multiple MRI modalities is through Monte-Carlo (MC) simulations in realistic tissue phantoms, especially given the advancements in volume microscopy and machine learning for imaging and digitally segmenting large mesoscale tissue samples. Historically, MC simulations in MRI have been used since the 1990s[37,38] in both artificial substrates and more realistic microscopy images. Over the past two decades, several synthetic media generation methods[11,37,47–56,39,57–59,40–46] have been presented in both white matter (WM) and gray matter (GM), with larger microscopy samples used to validate and investigate diffusion MRI. While the literature covers simulations investigating internal magnetic fields, current studies are limited to using 2D images or considering only a few axons[11,37,49–51,60–63]. Hence, microstructural information across multiple modalities on the mesoscopic scale has so far not been extensively investigated.

In this study we take another step towards investigating the effect of internal magnetic fields on the MRI signal by using publicly available mesoscale volume microscopy data[64,65] of rat brain white matter axons. The data, comprising myelinated axon segments from traumatic brain injury (TBI) and SHAM rats, allow us to simulate over 10,000 realistically organized axons simultaneously. Assuming scalar susceptibility, we calculate the magnetic field distribution induced by these axons and perform Monte Carlo simulations to compute MGE signals, asymmetric spin echo (ASE) signals, and pulsed-gradient spin echo (PGSE) signals.

Our objectives are twofold:

1. **Objective 1 (O1):** To determine if MRI can estimate the mesoscopically averaged magnetic field via the phase of the MGE and ASE signals.
2. **Objective 2 (O2):** To evaluate whether our analytical model of cylinders with arbitrary orientation distribution can predict the mesoscopically averaged magnetic field when combined with fODF information estimated with diffusion MRI.

Our work represents an important step towards validating the use of dMRI information to account for sub-voxel frequency shifts caused by the structural anisotropy of WM, which is crucial for robust susceptibility estimation.

# 3| Methods

Here we give an overview of the WM substrates used to study the effect of the induced magnetic fields $\Delta \boldsymbol{B}(\boldsymbol{r})$ from WM axons on the MRI signal, and the MC simulation used to generate MRI signals with and without diffusion weighting. All simulations were done in MATLAB (The Mathworks, Natick, MA, USA) on a desktop PC equipped with a dedicated Nvidia RTX 4070 Ti 12 GB GPU, a 13$^{\text{th}}$ generation 16 core Intel i7 processor customized with 192 Gb of DDR5-5200MHz RAM. The code used to process the data, calculate the magnetic fields, and perform MC simulations is available upon reasonable request to the corresponding author.

### Processing WM Substrate
We used openly available data of segmented volume electron microscopy (EM) of rat brain WM[64–66]. We refer to previous studies for a more detailed description of the WM substrates.

The data included mesoscopically sized samples of WM with a resolution of 50 nm$^3$. The WM includes corpus callosum (CC) and cingulum bundle (CG) from SHAM rats, and rats undergone traumatic brain injury (TBI). Here we considered eight segmented regions of WM from two SHAM rats (SHAM-25 and SHAM-49) and two TBI rats (TBI-24 and TBI-2), from either the contra-lateral (contra) or ipsilateral (ipsi) side, where the labels are kept in correspondence with the original data. The available MATLAB data contained fully segmented and labelled intra-axonal spaces, and a binary mask of the myelin sheath. Figure 1 gives an overview of the 8 WM substrates.

Prior to calculating the magnetic field and performing MC simulations in these substrates, we made the following refinements to the axons: First, we down sampled the resolution to 0.1 µm$^3$. This was necessary for feasible calculation of $\Delta \boldsymbol{B}(\boldsymbol{r})$ across the whole sample. The matrix size of the voxelized grid after down-sampling is listed in Table 1. Second, we nulled all voxels in the intra-axonal space that touched a neighboring axon's intra-axonal space to make sure our MC simulation did not "leak" spins across axons. Third, we performed the following operations iteratively for the intra-axonal space of each segmented axon: We morphologically closed the intra-axonal space to remove any small spurious errors on the surface. Next, we removed all holes in the intra-axonal space, e.g., from segmented cells inside the axons, and checked that each axon consisted of one connected intra-axonal space. If an axon consisted of multiple pieces, we kept only the largest piece. If an axon exhibited branching, the axon was not used for MC, as such typically originated from errors during segmentation. We then computed the center of mass (COM) points for the cross-sectional planes perpendicular to an axon's main direction. Highly curved axons, which constituted less than 1% of the microstructure, were discarded to simplify the MC simulation. If the axons, for example, were pointing mainly along *z*, we stored a directional label 3, its direction $\hat{\boldsymbol{n}}$, and the COM points computed in the *xy* plane for each *z*-position. We then checked and removed any touching voxels from different axons. Axons pointing along *z* were truncated 5 µm above their bottom end due to a visually noticeable systematic curvature that we suspect originated from cutting the tissue sample. If an axon close to the volume boundary touched the side walls of the sample volume, we truncated it such that the whole axon never touched any boundary within the volume and kept only the largest segment. We prepared the ends of the axons such that we could create fully periodic boundaries in our MC simulation, inside the axon, which was crucial to not introduce additional orientation dispersion. This was done by finding two similar (Dice similarity index above 0.9) cross-sectional planes perpendicular to the axon's main direction close to the ends of the axon. We then fused the ends together such that their cross-section was the same. We stored the COM positions of the two ends which were to be used for MC simulations. After all these operations, we calculated the length based on the line following the COM points. Only axons longer than 20 µm was used for MC. Using the toolbox Rawseg[67,68], we calculated the EM-related fODF $\mathcal{P}^{\text{EM}}(\hat{\boldsymbol{n}})$ and its expansion coefficients $p_{2m}^{\text{EM}}$ in spherical harmonics $Y_l^m(\hat{\boldsymbol{n}})$

$$\mathcal{P}^{\text{EM}}(\hat{\boldsymbol{n}}) \approx 1 + \sum_{l=2,4,\ldots}^{l_{max}} \sum_{m=-l}^{l} p_{2m}^{\text{EM}} Y_l^m(\hat{\boldsymbol{n}}). \tag{1}$$

To simulate the microstructure coarse-grained by diffusion, Rawseg calculates the axon's COM tangent lines, after smoothing each COM line by a Gaussian filter of size $L = \sqrt{6D_0\Delta}$ for different diffusion times of Δ =10, 40, 70 and 100 ms. We therefore denote $p_{2m}^{\text{EM}}(\Delta)$ as a function of Δ. Lastly, we checked for overlapping voxels one last time, and made sure that the ends remained identical, when considering all the processed axons in the whole grid. The intra-axonal voxels defined an indicator function $v_a(r)$ where $r$ denotes grid positions. The resulting intra-axonal volume fractions $\zeta_a$ for each substrate are listed in Table 1.

|  | **Ipsilateral** | | | | **Contralateral** | | | |
| --- | --- | --- | --- | --- | --- | --- | --- | --- |
|  | $\zeta_a$ | $\zeta_m$ | DIM | #MC Axons | $\zeta_a$ | $\zeta_m$ | DIM | #MC Axons |
| **SHAM-25** | 0.12 | 0.49 | 1006 × 2035 × 685 | 14804 | 0.18 | 0.49 | 2028 × 1001 × 646 | 20594 |
| **SHAM-49** | 0.18 | 0.48 | 1003 × 1006 × 598 | 8496 | 0.13 | 0.45 | 1006 × 2033 × 493 | 11607 |
| **TBI-24** | 0.10 | 0.45 | 1061 × 1453 × 618 | 10850 | 0.10 | 0.50 | 997 × 2024 × 638 | 14757 |
| **TBI-2** | 0.07 | 0.43 | 1005 × 2034 × 303 | 7760 | 0.11 | 0.47 | 1008 × 2028 × 536 | 12831 |

Table 1: $\zeta_a$ is the intra-axonal volume fraction of the axons used for Monte-Carlo simulations, $\zeta_m$ is the volume fraction of the myelin sheath used for generating the induced magnetic field. DIM indicates the matrix size of each substrate while #MC Axons indicate the number of axons and not the actual intra-axonal volume fraction of substrates.

The original segmented myelin mask was also morphologically closed and small objects were removed. Next, the myelin mask was multiplied by a negated mask of the processed intra-axonal mask, to remove any overlapping voxels. The myelin mask describing the indicator function $v_m(r)$ contained all the original axon's myelin with volume fraction $\zeta_m$ as listed in Table 1, and this myelin mask was used to calculate the field distribution, while the labelled grid $v_A(r)$ of intra-axonal spaces of axons longer than 20 μm was used for MC simulations.

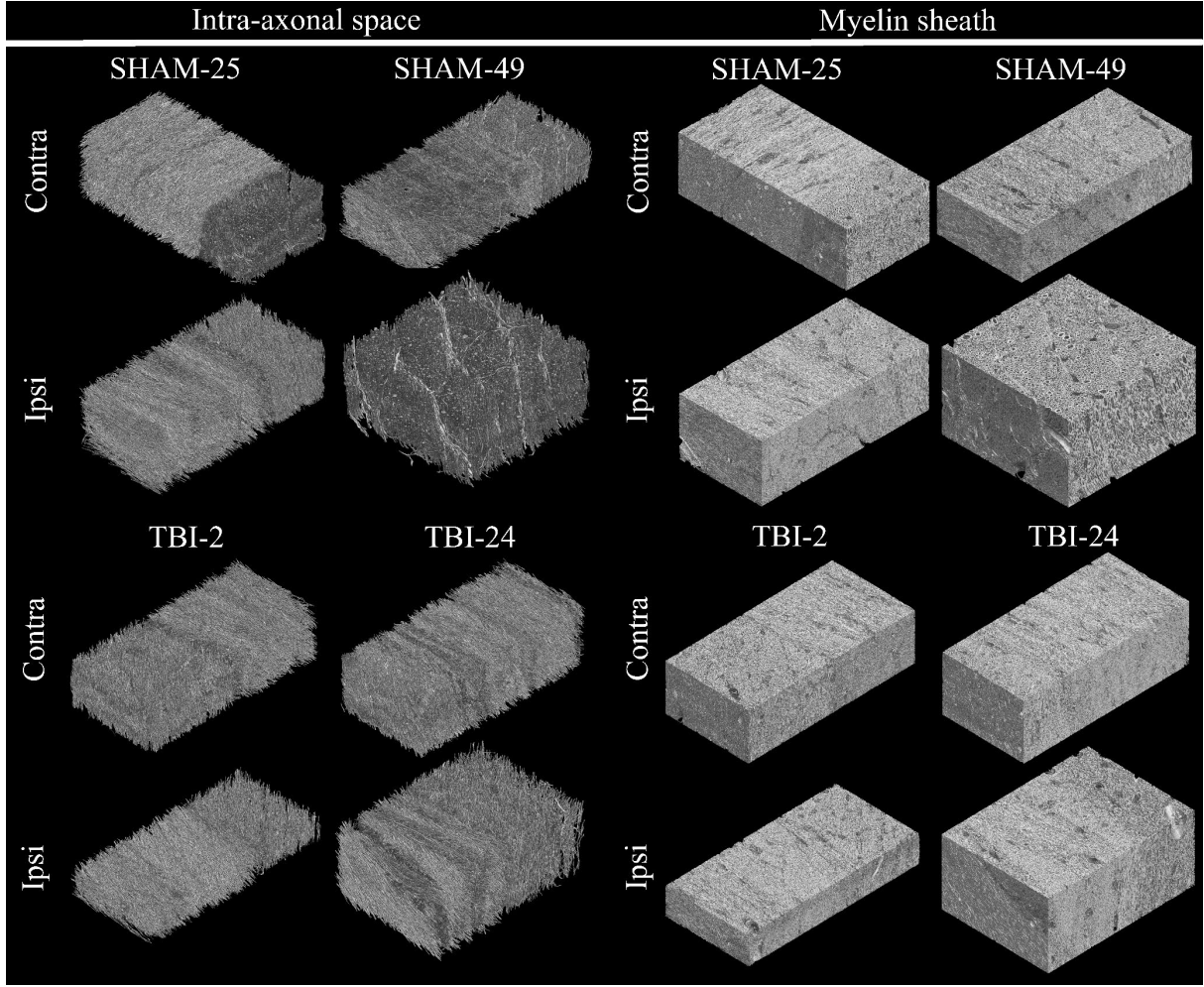

*Figure 1 - In-silico white matter axon phantoms used for Monte-Carlo simulations. Eight different substrates were used for Monte-Carlo simulations from two different SHAM rats labelled 25 and 49 and two different TBI rats labelled 2 and 24. Labels correspond to the original data. For each brain, both ipsilateral (ipsi) and contralateral (contra) tissue samples are considered. The tissue is extracted from the corpus callosum and cingulum bundle. The intra-axonal spaces are used for the Monte-Carlo simulation of diffusing spins, while the myelin sheaths constitute the field-inducing tissue, perturbing the Larmor frequency of the diffusing spins.*

**Larmor Frequency shift $\Omega(\boldsymbol{r})$**

For our MC simulation we considered signal dephasing due to the (demeaned) Larmor frequency shift $\Omega(\boldsymbol{r})$ induced by the voxelized susceptibility-weighted myelin grid $\chi_m v_m(\boldsymbol{r})$

$$\begin{aligned}\Omega(\boldsymbol{r}) &\cong \gamma\, \hat{\mathbf{B}}^T \Delta \boldsymbol{B}(\boldsymbol{r}) = \gamma B_0\, \hat{\mathbf{B}}^T \int_V d\boldsymbol{r}'\, \boldsymbol{\Upsilon}(\boldsymbol{r}-\boldsymbol{r}')\delta\chi_m(\boldsymbol{r}')\hat{\mathbf{B}} \\ &= \gamma B_0\, \hat{\mathbf{B}}^T[\delta\chi\otimes\boldsymbol{\Upsilon}](\boldsymbol{r})\hat{\mathbf{B}}.\end{aligned} \quad (2)$$

Here $\boldsymbol{\Upsilon}(\boldsymbol{r})$ is the dipole field tensor, $\delta\chi_m(\boldsymbol{r}) = \chi_m v_m(\boldsymbol{r}) - \overline{\chi}$ is the zero-mean magnetic susceptibility of the axons, which is computed such that the average Larmor frequency shift across the whole volume V is zero, and $\overline{\chi} = \zeta_m \chi_m$ denotes the mesoscopically averaged magnetic susceptibility. The reason for subtracting the frequency shift from the bulk susceptibility $\overline{\chi}$ is a matter of convention and is done in order to compare the average frequency shift inside the intra-axonal space $\overline{\Omega}_a$ to our previously derived analytical result[14] for the mesoscopic averaged Larmor frequency shift $\overline{\Omega}^{\text{Meso}}$ in a mesoscopic region with zero-mean frequency shift (the frequency shift from $\overline{\chi}$ are instead captured by the so-called macroscopic frequency shift $\overline{\Omega}^{\text{Macro}}$, which is the only shift considered in QSM). We used an intrinsic scalar susceptibility of $\chi_m = -100$ ppb/$\zeta_m$ such that $\overline{\chi} = -100$ ppb agreed with experiments[14,69,70]. We neglected microscopic susceptibility anisotropy of myelin, firstly because evidence suggests it contributes less than structural anisotropy of axons[70], and secondly because the EM resolution was inadequate for a faithful simulation. The sample averaged frequency shift $\overline{\Omega}_a$ inside the axons is

$$\overline{\Omega}_a(\mathbf{B}_0) = \frac{1}{V\zeta_a} \int_V d\mathbf{r}\, v_a(\mathbf{r})\Omega(\mathbf{r}). \tag{3}$$

Here $\Omega(\mathbf{r})$ and $\overline{\Omega}_a(\mathbf{B}_0)$ were computed numerically for the 3D voxelized microstructure in an external magnetic field $\mathbf{B}_0 = B_0\hat{\mathbf{B}}$, with $B_0 = 3$ T or 7T and 13 different orientations $\hat{\mathbf{B}}$ found using electrostatic repulsion.[71] We thus evaluated our MC simulated MR signal for 26 different frequency distributions in total. For later convenience in our MC simulation, we utilized that every distribution $\Omega(\mathbf{r})$ consists of 6 independent tensor components $\mathbf{A}(\mathbf{r}) \equiv [\delta\chi \otimes \mathbf{Y}](\mathbf{r})$, containing the information of the induced magnetic field from the microstructure, which can be computed using the convolution theorem and fast Fourier transform, such that $\Omega(\mathbf{r}) = \gamma B_0 \hat{\mathbf{B}}^T \mathbf{A}(\mathbf{r})\hat{\mathbf{B}}$.

**Monte-Carlo simulation**
Here we give an overview of the MC simulation performed in the WM substrates. Since this required doing MC simulations in many axons at once, it was crucial that the step and collision type were chosen to make the random walk as computationally simple as possible. Each particle's random step was drawn from a uniform distribution with step length $\delta_l = 0.1$ μm equal to the voxel resolution. The intrinsic diffusivity was set to $D_0 = 2$ μm$^2$/ms, leading to a time step $\delta_t = 0.83$ μs. When a random walker stepped outside its intra-axonal space, the step was rejected. This rejection sampling scheme led to approximately a 6% reduction in the intra-axonal axial diffusivity (see Supporting Material), which is acceptable since the orientation distribution is the focus of this study and not the diffusivity. We used on the order of N ~ $10^7$ particles, corresponding to a particle density of 72 particles/μm$^3$ inside the axons. While this is inarguably too few to study diffusion inside individual axons, it was found to robustly estimate the signal from the entire population of axons (see Supporting Material). Performing the MC simulation and computing all the signals for a diffusion time around $\Delta = 100$ ms took around 20 hours per substrate.

*MC Signal Generation*
The MC simulations were used to simulate a multi gradient echo signal $S_{\text{MGE}}$ (MGE), an asymmetric spin echo (ASE) signal $S_{\text{ASE}}$, and a spin-echo (SE) signal with diffusion weighting $S_{\text{PGSE}}$. Figure 2 gives an overview of the encoded signals. To include signal dephasing caused by the varying Larmor frequency $\Omega(\mathbf{r})$, we only had to accumulate a normalized signal tensor $\boldsymbol{\varphi}(T_E; p, t)$ for each random walker $p$, corresponding to each of the six independent tensor elements in $\mathbf{A}$ (as described in the previous section), and for each encoded echo time $T_E$,

$$\boldsymbol{\varphi}(T_E; p, t) = \alpha_{T_E}(t)\boldsymbol{\varphi}(p, t-1) + \delta_t \mathbf{A}\left(\boldsymbol{r}_p(t)\right). \tag{4}$$

Here the function

$$\alpha_{T_E}(t) = \begin{cases} 1, & t \neq \dfrac{T_E}{2} \\ -1, & t = \dfrac{T_E}{2} \end{cases} \tag{5}$$

specifies when a 180-degree RF pulse (here an ideal phase flip) occurs and $\boldsymbol{r}_p(t) = \boldsymbol{r}_{p0} + \boldsymbol{\delta r_p}(t)$ denotes the position of the particle, where $\boldsymbol{r}_{p0}$ is the initial position and $\boldsymbol{\delta r_p}(t)$ the displacement. Numerically, $\boldsymbol{\varphi}$ was stored as an array with dimensions given by the total number of particles, number of echo times $T_E$ and the 6 tensor components of $\mathbf{A}$. For the MGE signal, $\alpha(t) = 1$ for all $t$ and here we write the normalized signal tensor as $\boldsymbol{\varphi}(p; t)$ for simplicity. The ASE signal $S_{\text{ASE}}(T_E + \Delta T_E)$ at a time $T_E + \Delta T_E$, where $\Delta T_E$ denotes the asymmetric delay after $T_E$, were computed for all combinations $\gamma B_0 \widehat{\mathbf{B}}^T \boldsymbol{\varphi}(T_E + \Delta T_E; p)\widehat{\mathbf{B}}$ of $\widehat{\mathbf{B}}$ and $B_0$

$$S_{\text{ASE}}(T_E + \Delta T_E) = \frac{1}{N}\sum_p e^{-\gamma B_0 \widehat{\mathbf{B}}^T \boldsymbol{\varphi}(T_E+\Delta T_E;p)\widehat{\mathbf{B}}} \quad \text{(ASE)}, \tag{6}$$

$$S_{\text{MGE}}(t) = \frac{1}{N}\sum_p e^{-\gamma B_0 \widehat{\mathbf{B}}^T \boldsymbol{\varphi}(p;t)\widehat{\mathbf{B}}} \quad \text{(MGE)}.$$

The MGE signal was calculated up to $t = 40$ ms, while the ASE signal was calculated at $T_E = 80$ ms with $\Delta T_E$ up to 20 ms.

*Pulsed-Gradient diffusion signal $S_{PGSE}$*
We computed the PGSE signal $S_{\text{PGSE}}$ by adding a diffusion gradient $\mathbf{g}$ to the SE signal encoding, see Figure 2 for an overview. The gradient $\mathbf{g}$ was chosen such that $b = 0, 1, 3, 4, 5, 7, 10$ ms/µm² and with $\hat{\mathbf{g}}$ along 30 gradient directions for $b < 5$ ms/µm², and 60 directions for the remaining. The directions were selected using electrostatic repulsion[63]. The diffusion time was either $\Delta = 10, 40, 70, 100$ ms, while the gradient pulse duration was equal to the time step, $\delta = \delta_t$. The echo time considered for the PGSE signal was $T_E = 15\text{ms} + \Delta$ to minimize computation time. Then, to compute the signal $S(b, \hat{\mathbf{g}})$, we first compute the contribution to particle p's signal phase from the diffusion weighting as $\gamma\delta_t\sqrt{b/(\Delta - \delta/3)}\hat{\mathbf{g}} \cdot \Delta \mathbf{l}_p$, where $\Delta \mathbf{l}_p = \left(\boldsymbol{\delta r_p}(5 \text{ ms} + \Delta) - \boldsymbol{\delta r_p}(5 \text{ ms})\right)$ is the spatial displacement vector of particle p, and then calculate the sum

$$S_{\text{PGSE}}(b, \hat{\mathbf{g}}) = \frac{1}{N}\sum_p e^{-i\gamma B_0 \widehat{\mathbf{B}}^T \boldsymbol{\varphi}(T_E;p)\widehat{\mathbf{B}} - i\gamma\delta_t\sqrt{b/(\Delta - \delta/3)}\hat{\mathbf{g}}\cdot\Delta\mathbf{l}_p} \quad \text{(PGSE)}. \tag{7}$$

Notice $S_{\text{PGSE}}(0) = S_{\text{ASE}}(T_E)$ when $b = 0$.

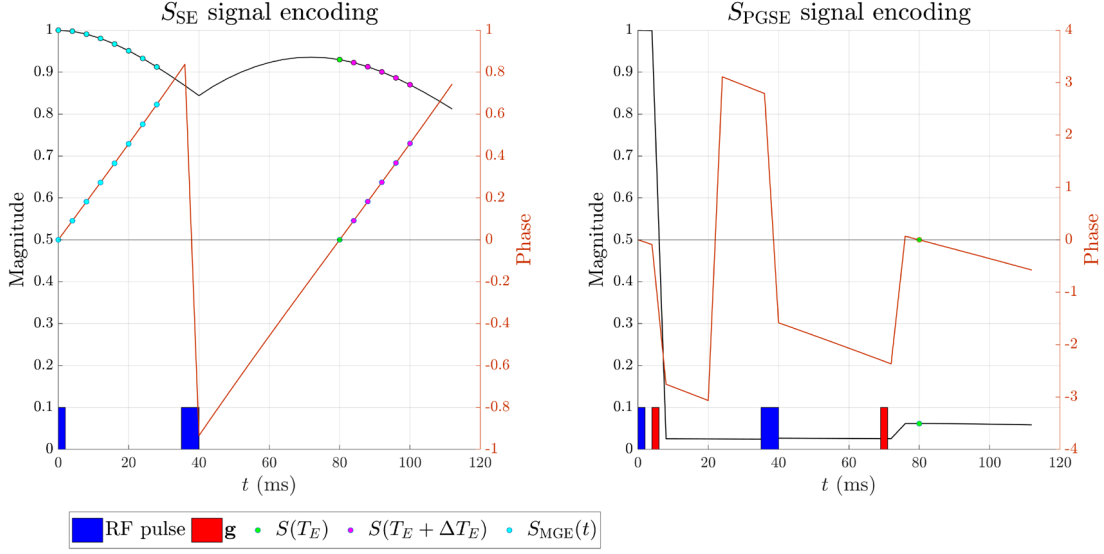

*Figure 2 – Signal encoding for SHAM-49-ipsi. Left figure shows encoding of spin echo signal. Blue tiles indicate the 90-degree and 180-degree RF pulse. Cyan point shows the measurement of the gradient echo signal, here up to $t = 40\ ms$. Green point show spin echo signal at the echo time, here $T_E = 80\ ms$, while magenta points indicate measuring the spin echo signal at a later time, here up to $\Delta T_E = 20\ ms$. Right figure shows encoding of PGSE signal with b=10 ms/µm² and gradient direction roughly perpendicular to main fiber orientations. Signal attenuation at the first gradient pulse is roughly $sinc(qL/2)$, $\sim 10^{-2}$ as $q \approx 1/6$ and $L \sim 100$ µm corresponding to the variance of particle positions after 5 ms, where the first gradient pulse was turned on. Red tiles indicate the pulsed gradients.*

## Parameter estimation

### The Standard Model of diffusion in WM

For long diffusion times, the signal from a single non-exchanging axon with orientation $\hat{\boldsymbol{n}}$ is typically modelled by a gaussian stick compartment[2] (zero radial diffusivity). The signal kernel from a bundle of parallel sticks is thus

$$\mathcal{K}(b, \hat{\boldsymbol{n}} \cdot \hat{\boldsymbol{g}}) = e^{-bD_a(\hat{\boldsymbol{n}} \cdot \hat{\boldsymbol{g}})^2}, \text{(Stick compartment)}. \tag{8}$$

For a collection of non-exchanging stick bundles with different orientations, the total signal is obtained by a convolution with the fODF $\mathcal{P}^D(\hat{\boldsymbol{n}})$, where the superscript '$D$' for "diffusion" is to distinguish it from the EM-derived fODF, $\mathcal{P}^{EM}$:

$$S_{SM}(b, \hat{\boldsymbol{g}}) = S_0 \int d\hat{\boldsymbol{n}}\, \mathcal{P}^D(\hat{\boldsymbol{n}})\mathcal{K}(b, \hat{\boldsymbol{n}} \cdot \hat{\boldsymbol{g}}). \tag{9}$$

The fODF $\mathcal{P}^D(\hat{\boldsymbol{n}})$ can also be expanded in spherical harmonics $Y_l^m(\hat{\boldsymbol{n}})$

$$\mathcal{P}^D(\hat{\boldsymbol{n}}) \approx 1 + \sum_{l=2,4,\ldots}^{l_{max}} \sum_{m=-l}^{l} p_{lm}^D Y_l^m(\hat{\boldsymbol{n}}), \tag{10}$$

where $p_{lm}^D$ are the diffusion-derived fODF expansion coefficients. In practice, the apparent fODF and thus the $p_{lm}^D$ depend on the diffusion time[67], and so we write $p_{lm}^D(\Delta)$.

We also consider $S_{SM}(b,\hat{\mathbf{g}})$ with an intra-axonal axial kurtosis[72] $W_a^{\parallel}$, such that the kernel becomes

$$\mathcal{K}(b,\hat{\mathbf{n}}\cdot\hat{\mathbf{g}}) = e^{-bD_a(\hat{\mathbf{n}}\cdot\hat{\mathbf{g}})^2 + \frac{1}{6}(bD_a)^2(\hat{\mathbf{n}}\cdot\hat{\mathbf{g}})^4 W_a^{\parallel}}, \text{ (Stick with axial kurtosis)}. \qquad (11)$$

Upon fitting $S_{SM}(b,\hat{\mathbf{g}})$ to the MC simulated PGSE signal $S_{PGSE}$, we computed the difference in the Bayesian information criterion[73] $\Delta BIC$ to check the change in residual variance compared to adding an additional parameter. A difference of $\Delta BIC < -6$ would support a non-vanishing $W_a^{\parallel}$.[74]

*Mesoscopic Larmor frequency shift from cylinders with arbitrary orientation distribution*
One of the main goals of this study was to compare $\overline{\Omega}_a(\mathbf{B}_0)$ with our previously derived analytical result[14]

$$\overline{\Omega}^{\text{Meso}}(\mathbf{B}_0) = -\gamma B_0 \bar{\chi}\,\frac{1}{2}\left(\mathbf{T}(\hat{\mathbf{B}}) - \frac{1}{3}\right) \qquad (12)$$

for the mesoscopically averaged Larmor frequency shift $\overline{\Omega}^{\text{Meso}}$ for long multi-layered cylinders with scalar susceptibility and arbitrary orientation dispersion. Equation (12) is valid for both the average intra-axonal and average extra-axonal frequency shifts (the latter not considered here). Here $\mathbf{T}(\hat{\mathbf{B}}) = \hat{\mathbf{B}}^{\mathrm{T}}\mathbf{T}\hat{\mathbf{B}}$ where $\mathbf{T}$ is the fODF scatter matrix[35], which can be written in terms of the $l=2$ expansion coefficients $p_{2m}^{\Omega}$ of the ODF in spherical harmonics $Y_2^m$

$$\mathbf{T}(\hat{\mathbf{B}}) = \hat{\mathbf{B}}^{\mathrm{T}}\mathbf{T}\hat{\mathbf{B}} = \frac{1}{3} + \frac{2}{15}\sum_{m=-2}^{2} p_{2m}^{\Omega} Y_2^m(\hat{\mathbf{B}}). \qquad (13)$$

Notice we write $p_{2m}^{\Omega}$ to distinguish it from $p_{2m}^{EM}(\Delta)$ and $p_{2m}^{D}(\Delta)$, and is independent of diffusion time as $\overline{\Omega}^{Meso}$ describes the mean magnetic field[13].

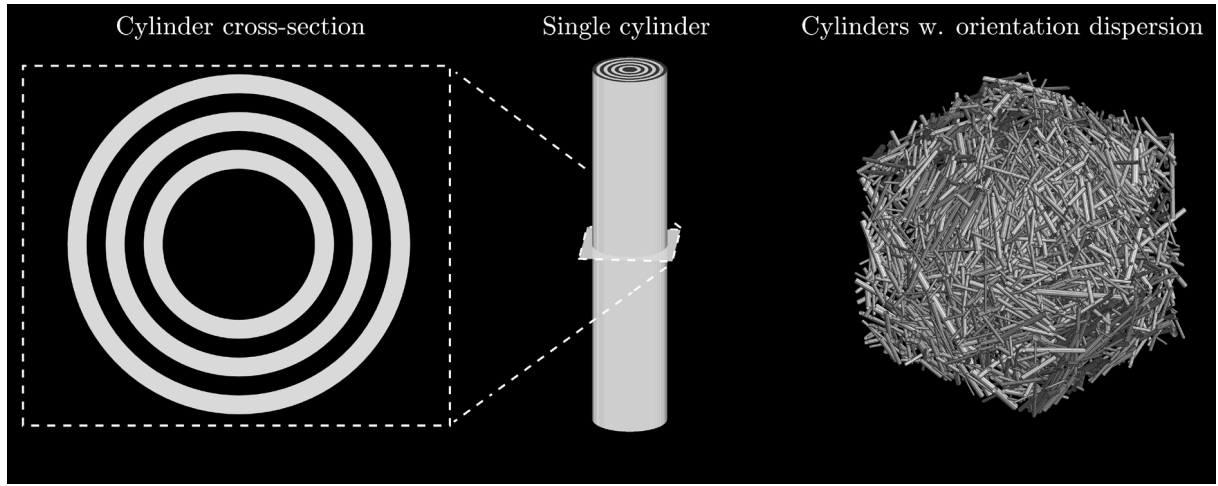

*Figure 3 – Biophysical model of WM axons as long hollow cylinders randomly positioned and with arbitrary orientation dispersion. Cylinders are assumed to have a scalar magnetic susceptibility $\chi_m$.*

## Analysis

Equation (12) was developed as a biophysical model of the measured frequency shift from myelinated axons based on infinitely long cylinders (see Figure 3. However, it remains to be validated. Since we cannot estimate $p_{2m}^{\Omega}$ solely from MGE data, where $\overline{\chi}$ is unknown, a central question is if $p_{2m}^{D}(\Delta) \approx p_{2m}^{\Omega}$ to a sufficient degree to accurately estimate $\overline{\Omega}_a(\mathbf{B}_0)/\overline{\chi}$ from $\overline{\Omega}^{Meso}(\mathbf{B}_0; \Delta)/\overline{\chi}$ and $p_{2m}^{D}(\Delta)$. This, in turn, would support the estimation of $\overline{\chi}$ for this particular microstructure with μQSM, an extension of QSM that incorporates sub-voxel frequency shifts from orientationally dispersed axons described by Eq. (12), and where the $p_{2m}^{D}(\Delta)$ are used to estimate $\mathbf{T}(\hat{\mathbf{B}})$ prior to susceptibility fitting. We therefore performed the following comparisons:

### O1)

We investigated at what gradient echo times the slope of the simulated MGE signal $S_{MGE}(t)$ phase agreed with the intra-axonal frequency shift $\overline{\Omega}_a(\mathbf{B}_0)$ generated by the induced magnetic field. This was to ensure that $\overline{\Omega}_a(\mathbf{B}_0)$ can be meaningfully measured from MR. We fitted the measured signal phase of $S_{MGE}(t)$ using time points up to a given $t_{max}$ up to 40 ms using either a linear or a 3$^{rd}$ order polynomial fit. The linear coefficient in either case was then compared to $\overline{\Omega}_a(\mathbf{B}_0)$ for varying $\hat{\mathbf{B}}$ and $B_0$. We also investigated the signal phase of $S_{ASE}(T_E + \Delta T_E)$, with $T_E = 40$ ms and $(\Delta T_E)_{max}$ up to 20 ms and performed the same comparison.

### O2)

Next step was to compare $\overline{\Omega}_a(\mathbf{B}_0)$ against $\overline{\Omega}^{\text{Meso}}(\mathbf{B}_0; \Delta)$ where $\mathbf{T}(\widehat{\mathbf{B}})$ was estimated using $\overline{\chi}$ and either $p_{2m}^{\text{D}}(\Delta)$ or $p_{2m}^{\text{EM}}(\Delta)$, i.e.

$$\overline{\Omega}_{\text{EM}}^{\text{Meso}}(\mathbf{B}_0; \Delta) = -\gamma B_0 \overline{\chi} \frac{1}{15} \sum_{m=-2}^{2} p_{2m}^{\text{EM}}(\Delta) Y_2^m(\widehat{\mathbf{B}}), \tag{14}$$

and

$$\overline{\Omega}_{\text{D}}^{\text{Meso}}(\mathbf{B}_0; \Delta) = -\gamma B_0 \overline{\chi} \frac{1}{15} \sum_{m=-2}^{2} p_{2m}^{\text{D}}(\Delta) Y_2^m(\widehat{\mathbf{B}}). \tag{15}$$

To test how well the fODF from EM and diffusion, respectively, could be used for estimating susceptibility, we computed the normalized root-mean-square-error (NRMSE) across $\widehat{\mathbf{B}}$ as a function of the diffusion time $\Delta$

$$\text{NRMSE}_{\text{EM}}(\Delta) = \frac{\sqrt{\frac{1}{N_{\widehat{\mathbf{B}}}} \Sigma_{\widehat{\mathbf{B}}} \left( \overline{\Omega}_{\text{EM}}^{\text{Meso}}(\mathbf{B}_0; \Delta) - \overline{\Omega}_a(\mathbf{B}_0) \right)^2}}{\max\left(\overline{\Omega}_a(\mathbf{B}_0)\right) - \min\left(\overline{\Omega}_a(\mathbf{B}_0)\right)} \tag{16}$$

and

$$\text{NRMSE}_{\text{D}}(\Delta) = \frac{\sqrt{\frac{1}{N_{\widehat{\mathbf{B}}}} \Sigma_{\widehat{\mathbf{B}}} \left( \overline{\Omega}_{\text{D}}^{\text{Meso}}(\mathbf{B}_0; \Delta) - \overline{\Omega}_a(\mathbf{B}_0) \right)^2}}{\max\left(\overline{\Omega}_a(\mathbf{B}_0)\right) - \min\left(\overline{\Omega}_a(\mathbf{B}_0)\right)} \tag{17}$$

Here we chose to normalize the RMSE to the range, $\max\left(\overline{\Omega}_a(\mathbf{B}_0)\right) - \min\left(\overline{\Omega}_a(\mathbf{B}_0)\right)$, since $\overline{\Omega}_a(\mathbf{B}_0)$ takes on both positive and negative values.

We also estimated the error ratio $\beta$ by minimizing the least squares difference across $\widehat{\mathbf{B}}$ as a function of the diffusion time $\Delta$

$$\beta_{\text{EM}}(\Delta) = \frac{1}{\overline{\chi}} \underset{\overline{\chi}'}{\text{argmin}} \frac{1}{2} \sum_{\widehat{\mathbf{B}}} \left( \overline{\Omega}_{\text{EM}}^{\text{Meso}}(\mathbf{B}_0; \Delta, \overline{\chi}') - \overline{\Omega}_a(\mathbf{B}_0) \right)^2, \tag{18}$$

and

$$\beta_{\text{D}}(\Delta) = \frac{1}{\overline{\chi}} \underset{\overline{\chi}'}{\text{argmin}} \frac{1}{2} \sum_{\widehat{\mathbf{B}}} \left( \overline{\Omega}_{\text{D}}^{\text{Meso}}(\mathbf{B}_0; \Delta, \overline{\chi}') - \overline{\Omega}_a(\mathbf{B}_0) \right)^2. \tag{19}$$

Hence, $\beta = 1$ corresponds to a perfect susceptibility fit.

# 4| Results

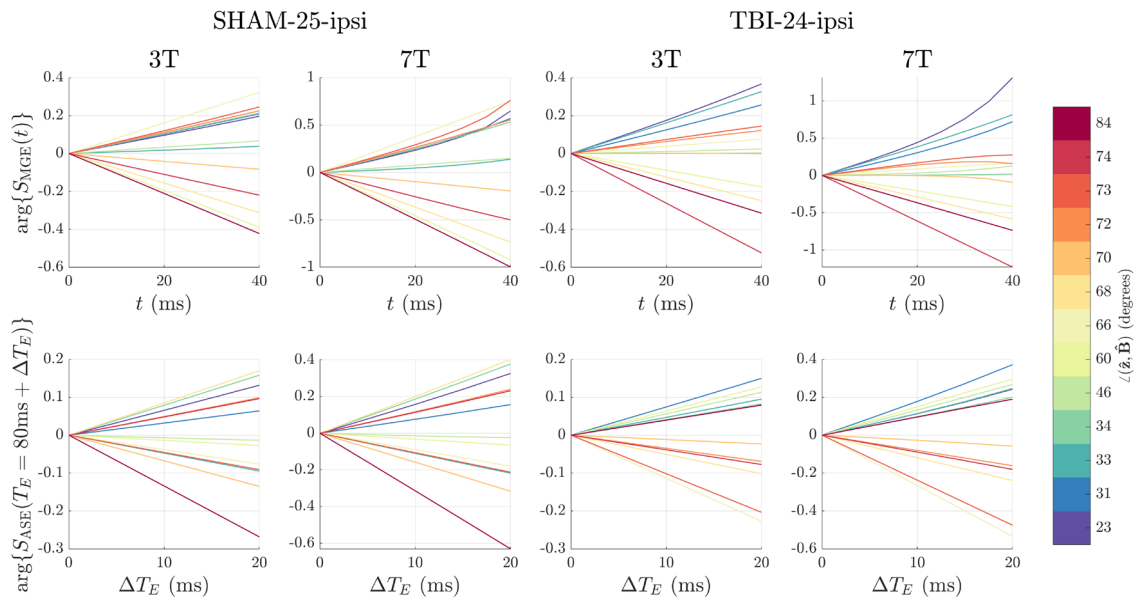

*Figure 4 – Upper row shows the signal phase, $arg\{S_{MGE}(t)\}$, of the MGE signal with an external field of either 3T or 7T for SHAM-25-ipsi and TBI-24-ipsi. Second row shows the phase $arg\{S_{ASE}(T_E + \Delta T_E)\}$ with $T_E = 80ms$. Colors corresponds to different angles between $\mathbf{B}_0$ and $\hat{\mathbf{z}}$.*

**O1)**

Figure 4 shows the MC signal phase for SHAM-25-ipsi of $S_{\text{MGE}}(t)$ and $S_{\text{ASE}}(T_E + \Delta T_E)$ for each field strength and direction. At 3T, very few directions exhibited non-linear behavior, which became more evident at 7T for $S_{\text{MGE}}(t)$. This shows that even though the signal comes from roughly 10.000 non-exchanging compartments and could therefore have been expected to contain higher-order cumulants proportional to the population variance of the Larmor frequency, the signal phase remained fairly linear. As described in a previous study[14], as time increases, and transverse relaxation in each compartment becomes more pronounced, the phase is well described by a power series in time, and the mean magnetic field may still be measurable by the linear term from a polynomial fit.

Figure 5 shows the mean ratio $\langle \overline{\Omega}_{\text{MGE}}(\mathbf{B}_0)/\overline{\Omega}_a(\mathbf{B}_0) \rangle_{\hat{\mathbf{B}}}$ and standard deviation (SD) between the linear term $\overline{\Omega}_{\text{MGE}}(\mathbf{B}_0)$ found from fitting the MC-simulated MGE signal phase to either a linear or 3$^{\text{rd}}$ order polynomial, and $\overline{\Omega}_a(\mathbf{B}_0)$. Both the mean and standard deviation are calculated over the $\hat{\mathbf{B}}$ directions, and the ratio is shown for different maximum times $t_{\text{max}}$ used in the fit. For 3T, linear fitting produced good results with a mean error $100\% \cdot (1 - \langle \overline{\Omega}_{\text{MGE}}(\mathbf{B}_0)/\overline{\Omega}_a(\mathbf{B}_0) \rangle_{\hat{\mathbf{B}}})$, and SD within 2% across all substrates, except TBI-24-ipsi that had an SD of 30%. The error was even further reduced by fitting a 3$^{\text{rd}}$ order polynomial, resulting in an SD of 5% for TBI-24-ipsi. At 7T, the 3$^{\text{rd}}$ order polynomial fitting again produced the best result. However, the mean error and standard deviation depended greatly on the microstructure. In SHAM-49-ipsi, which only contained cingulum, the mean error was still within 2% and SD around 4% for both linear and 3$^{rd}$ order fitting. Conversely, TBI-24-ipsi had an SD close to 110% for $t_{max} = 40$ ms. This was however reduced to an SD of 20% and mean of 7% by doing 3$^{rd}$ order fitting and using $t_{max} < 30$ ms.

We also considered a different approach of using instead the phase of the asymmetric spin echo $\overline{\Omega}_{ASE}$ at times $\Delta T_E$ after the echo $T_E$. Figure 6 shows the mean ratio $\langle \overline{\Omega}_{ASE}(\mathbf{B}_0)/\overline{\Omega}_a(\mathbf{B}_0) \rangle_{\hat{\mathbf{B}}}$ between the linear term $\overline{\Omega}_{ASE}(\mathbf{B}_0)$, found from fitting the MC simulated ASE signal phase after $T_E = 80$ ms to either a linear or 3$^{rd}$ order polynomial, and $\overline{\Omega}_a(\mathbf{B}_0)$. The ratio is shown for different maximum times $(\Delta T_E)_{max}$ included in the fit. Here we found the third order polynomial fitting to perform best, making the mean ratio less than 5% for all substrates. For TBI-24-ipsi the SD was reduced to 12%.

We also considered estimating the first cumulant including extra-axonal water. Figure 11 in supplementary material S1) shows for SHAM-ipsi-25, that it is equally possible to estimate the first cumulant when including extra-axonal water and that our theory remains accurate.

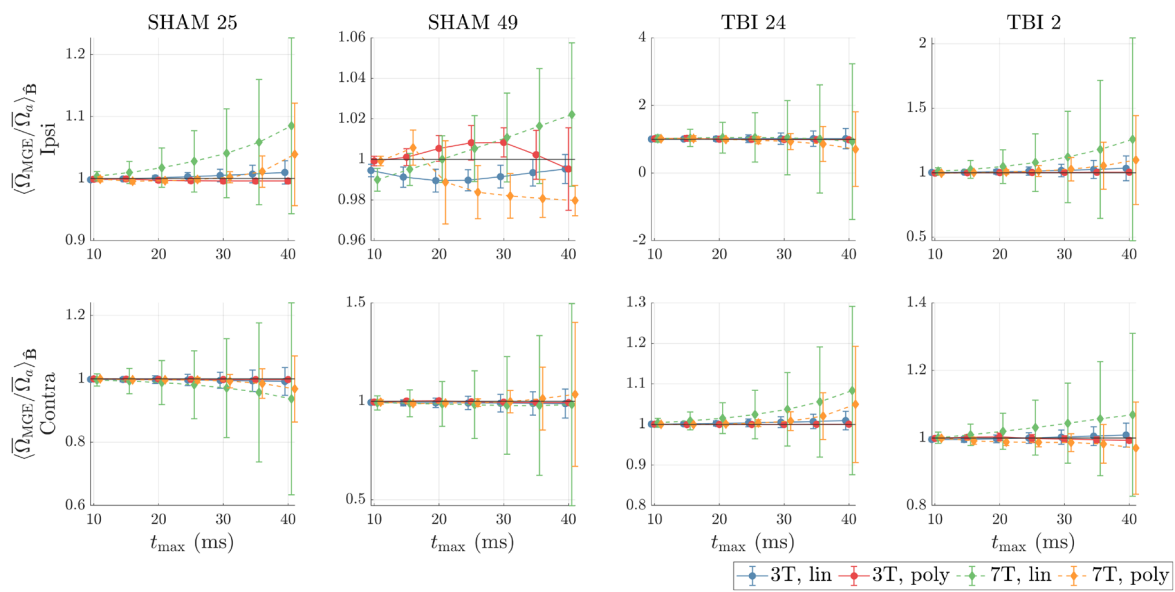

Figure 5 – Error ratio from fitting a linear or 3rd order polynomial to the phase of the MC simulated MGE signal $S_{MGE}$. The first row shows fitting ipsilateral substrates, while bottom row shows contralateral substrates. The rows correspond to different rat brains. Colors indicate different fitting protocols and field strengths. The x-axis indicates the maximum time used to fit the phase. (see legend).

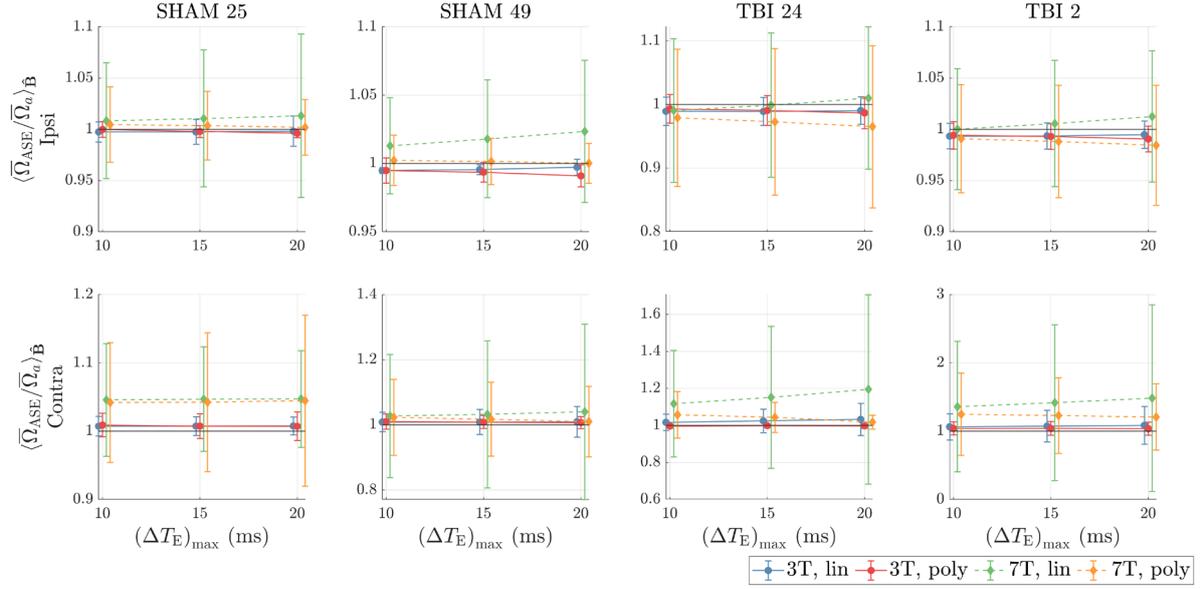

*Figure 6 – Error ratio from fitting a linear or 3rd order polynomial to the phase of the MC simulated SE signal $S_{SE}$ at times $\Delta T_E$ after the echo $T_E = 80\ ms$. The first row shows fitting ipsilateral substrates, while bottom row shows contralateral substrates. The rows correspond to different rat brains. Colors indicate different fitting protocols and field strengths. The x-axis indicates the maximum time used to fit the phase.*

**O2)**

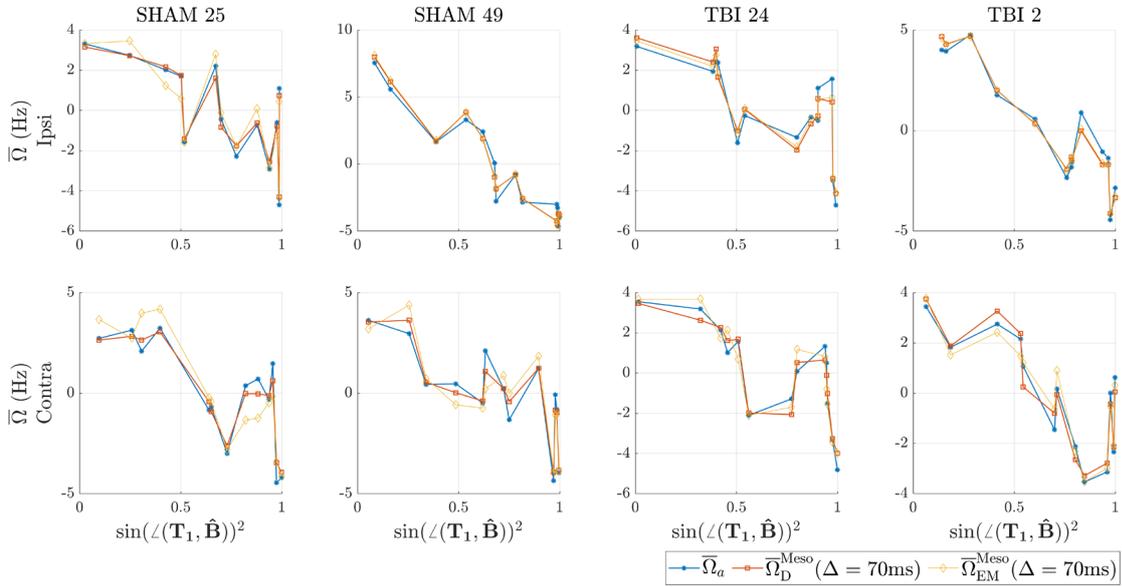

*Figure 7 – Larmor frequency shift for different WM substrates. Blue curve shows the intra-axonal frequency shift $\overline{\Omega}_a$ described by the mean magnetic field, while the red shows the estimated Larmor frequency shift $\overline{\Omega}_D^{Meso}$ based on the model for dispersed cylinders and estimated using the fODF estimated from MC simulated diffusion MRI signal with diffusion time $\Delta = 70\ ms$. The x-axis shows $\sin^2$ to the angle between the external field $\hat{B}$ and the principal eigenvector $T_1$ of the scatter matrix $T$.*

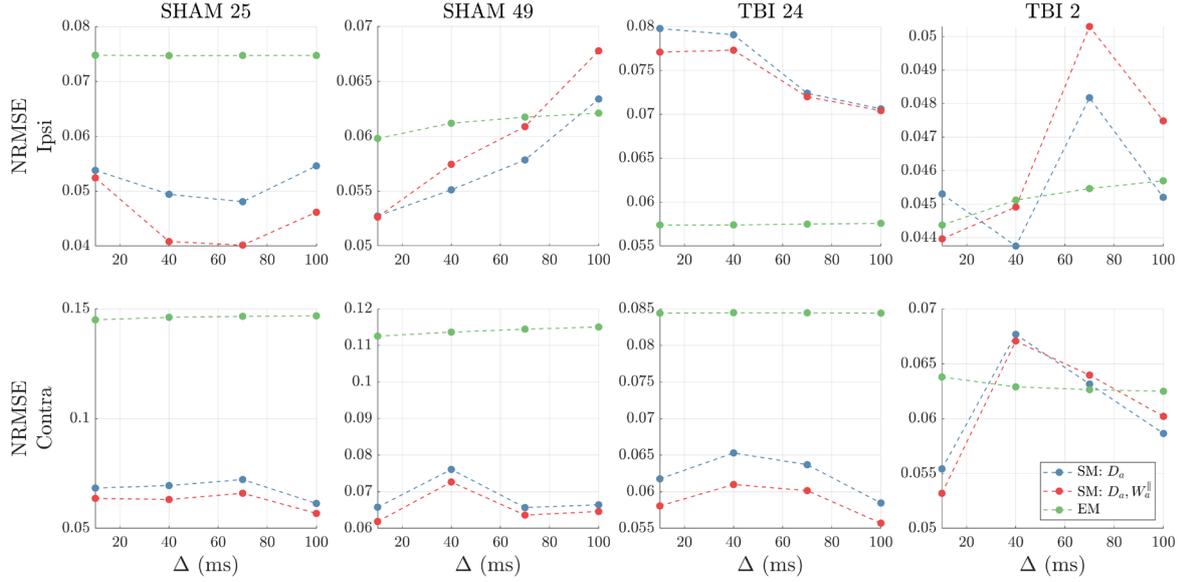

*Figure 8 – Normalized RMSE (NRMSE) between the simulated frequency shift $\overline{\Omega}_a$ and the theoretical mesoscopic frequency $\overline{\Omega}^{Meso}$ (cf. Eqs. (14)-(15)) estimated by using the fODF from fitting the standard model of diffusion to the simulated PGSE signal (blue and red) or using the fODF estimated directly from the EM data (in green). SM model fitting was done with (red) and without (blue) including intra-axonal axial kurtosis. The x-axis indicates the times Δ between the gradient pulses for the PGSE signal, or the time determining the size of the Gaussian filter ($\sqrt{2D_a\Delta}$) used to smooth the center of mass lines of each axon. Two first columns show for SHAM rats while the third and fourth column show for TBI rats.,*

Figure 7 shows $\overline{\Omega}_a(\mathbf{B}_0)$ and $\overline{\Omega}^{Meso}(\mathbf{B}_0; \Delta)$ across the different angles $\angle(\mathbf{T}_1, \widehat{\mathbf{B}}) = \mathrm{acos}(\mathbf{T}_1 \cdot \widehat{\mathbf{B}})$ between $\mathbf{B}_0$ and the principal eigenvector $\mathbf{T}_1$ of the estimated fODF scatter matrix. Here $B_0 = 7$ T.

Figure 8 shows the NRMSE (cf. Eqs. (16)-(17)) using the fODF estimated from EM or the fODF estimated from diffusion by fitting SM with or without axial kurtosis $W_a^\parallel$. In the figure, the internal fields were induced by an external field strength $B_0 = 7$T, but similar SM fitting results were found for 3T, and without internal fields, indicating a low bias in fODF estimation due to internal fields. For all substrates and across all diffusion times Δ, the NRMSE was around 5-8%, demonstrating a good correspondence between the first frequency shift $\overline{\Omega}_a(\mathbf{B}_0)$ and the model-based estimation $\overline{\Omega}^{Meso}(\mathbf{B}_0; \Delta)$ cf. Eq. (12). This is also evident from Figure 8.

Figure 9 shows the fitted β-values, $\beta_D(\Delta)$ and $\beta_{EM}(\Delta)$ (cf. Eqs. (18)-(19)) for different diffusion times. Here we found that $\beta_D(\Delta)$ was less than 10% for $\Delta \geq 70$ ms, across all substrates, if $W_a^\parallel$ was included. This shows that accounting for $W_a^\parallel$ for the intra-axonal signal may be important when estimating the fODF from diffusion MRI. Similar result were found when including extra-axonal water, which is shown for SHAM-ipsi-25 in Figure 12 in S1).

Figure 10 shows the estimated SM parameters along with the difference ΔBIC, from fitting all 8 substrates with and without $W_a^\parallel$. All parameters clearly change across all substrates, and the residual variance between the estimated and MC-simulated signal drops substantially when adding $W_a^\parallel$, which is reflected in a ΔBIC between -250 to -25 for the SHAM substrates and −500 to -200 for TBI substrates. Hence, the importance of accounting for $W_a^\parallel$ may depend on the axon morphology which changes due to TBI[66].

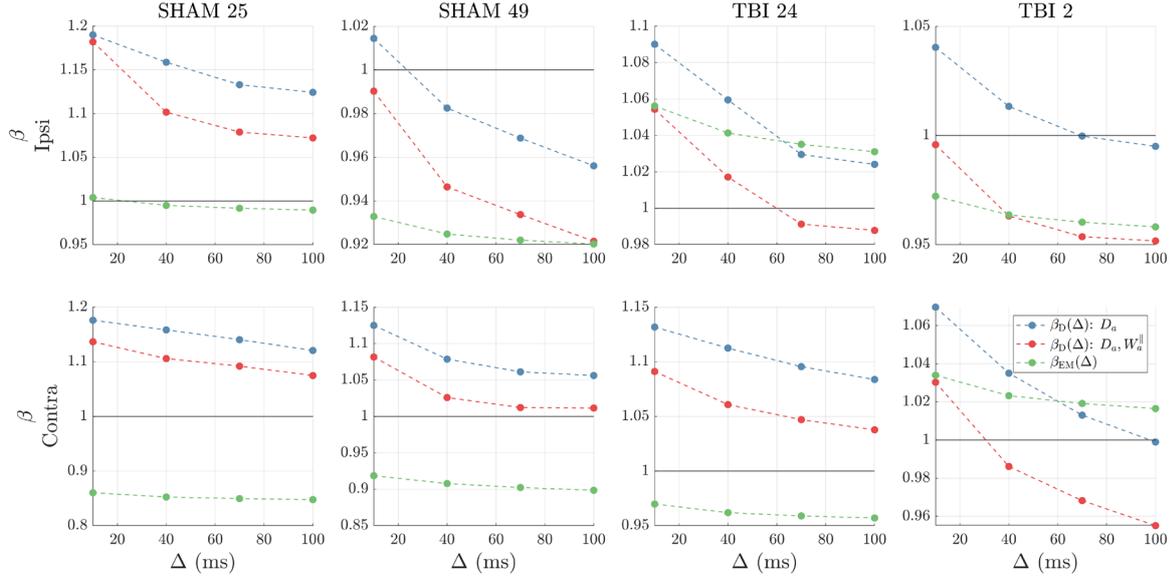

*Figure 9 – Normalized susceptibility fitting parameter β for all WM substrates. β was estimated by minimizing the squared difference between the frequency shift $\overline{\Omega}_a$ corresponding to the mean intra-axonal magnetic field and the estimated mesoscopic frequency $\overline{\Omega}^{Meso}$. The frequency was estimated using the fODF from fitting the standard model of diffusion to the simulated PGSE signal (blue and red) or using the fODF estimated directly from the EM data (in green). SM model fitting was done with and without intra-axonal axial kurtosis $W_a^{\parallel}$. The x-axis shows the time Δ controlling the width of the coarse graining filter and corresponds either to the time between the gradient pulses for the PGSE signal (for diffusion) or the size of the Gaussian filter used to smooth the center of mass lines of each axon (for EM).*

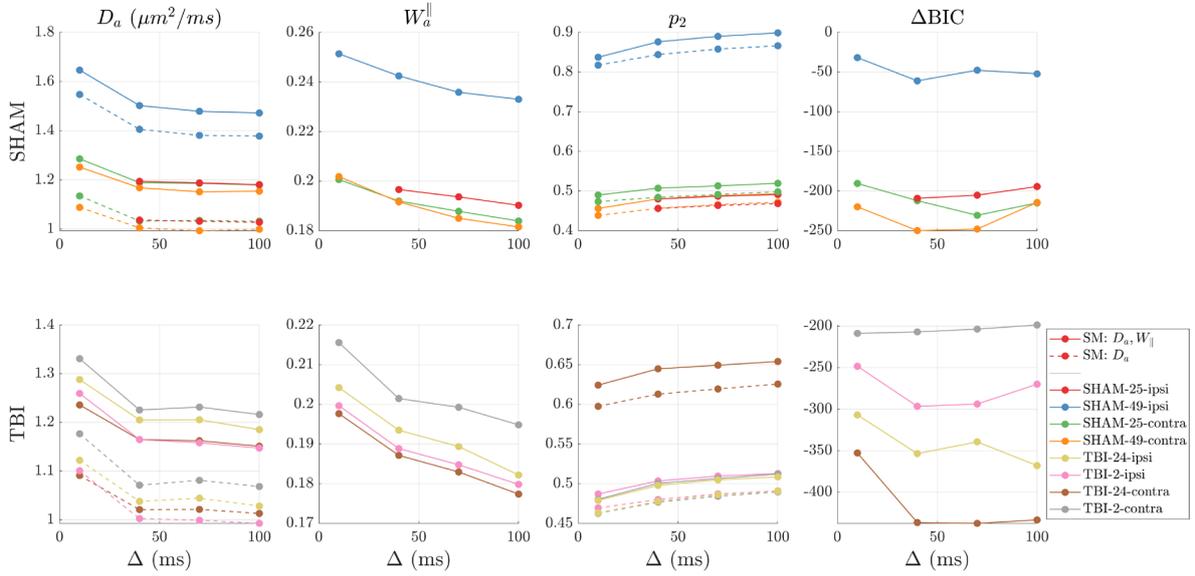

*Figure 10 – Fitted Standard model (SM) parameters for all eight substrates. The upper row shows parameters of SHAM rats and the lower row TBI rats. The colors correspond to the different substrates. Solid lines correspond to fitting SM with axial kurtosis, while dashed shows without (conventional SM). The upper row shows intra-axonal diffusivity $D_a$ and kurtosis $W_a^{\parallel}$, while the second row shows the dispersion parameter $p_2$ and the difference in Bayesian Information Criteria ΔBIC. We excluded the first red point at $\Delta = 10\ ms$ as the fit did not prove stable there, and the values became unrealistic.*

# 5 | Discussion

**Implications of realistic WM microstructure for the estimation of magnetic susceptibility**

In this work, we studied how the MRI signal depends on realistic WM axonal microstructure. This was done by performing Monte-Carlo (MC) simulations of the MRI signal for intra-axonal water in mesoscopically sized in-silico WM axon substrates from rat brain obtained from 3D electron microscopy[64,65]. Here we investigated if the orientation dependence of the multi gradient echo and asymmetric spin echo signal phases could be captured using information from diffusion MRI (dMRI). The connection to dMRI was recently described in our model framework, µQSM, which models the Larmor frequency shift from myelinated axons as orientationally dispersed concentric cylinders with scalar susceptibility.[14] In this model, the orientation dependence is captured by the *l=2* expansion coefficients $p_{2m}^{\Omega}$ of the cylinders' fiber orientation distribution function (fODF). Here we tested if the fODF estimated from dMRI could be used in our analytical solution for the mesoscopic frequency shift to predict the simulated signal phase. We therefore simulated a PGSE signal and fitted the Standard Model of Diffusion in White Mater (SM) to estimate $p_{2m}^{D}$ of the diffusion-derived fODF. Our results showed that incorporating $p_{2m}^{D}$ into µQSM could predict the measured mesoscopic frequency shift in the intra-axonal space from realistic WM axons with scalar susceptibility. This result was reproducible across four different substrates from SHAM rats and four different substrates from rats undergone Traumatic Brain Injury (TBI). Deviations between $p_{2m}^{D}$ and $p_{2m}^{\Omega}$ will inevitably arise since $p_{2m}^{D}$ was determined solely from the intra-axonal spaces of the axons used for MC, while $p_{2m}^{\Omega}$ comes from the mean shift inside the same intra-axonal spaces, but induced by all axons' myelin sheath – including the ones not used for MC. The reason for using all axon's myelin sheath was that the density of axons was not uniform over the whole volume, resulting in a non-uniform bulk susceptibility inside the sample. However, in µQSM, it is assumed that the bulk susceptibility is constant across the mesoscopic volume (voxel in practice), and this requirement was better fulfilled by retaining the realistic density of axons across the volume.

**Implications of realistic WM microstructure for the estimation of Standard Model parameters**

Previous studies have shown that SM parameters correlated with histological features of these axon substrates[66]. Our simulations also demonstrate that for realistic diffusion times, intra-axonal axial kurtosis affects the measured PGSE signal and may need to be accounted for to estimate the fODF. This was evident based on improvement in fODF estimation, non-negligible differences in SM parameters, and a lower Bayesian information criterion compared to having only an intra-axonal axial diffusivity. For example, including $W_a^{\parallel}$ increased the intra-axonal diffusivity $D_a$ around 10-20%. We also tried adding a non-zero radial diffusivity, but this could not explain enough signal variance to be justified based on the Bayesian information criterion. As described in previous studies[68], and reproduced here (see appendix), axon morphology such as caliber variation, tortuosity, etc., makes $W_a^{\parallel}$ decay as a power law, and for realistic diffusion times, it can produce a non-negligible effect on the diffusion signal. Based on our results, we therefore recommend to consider $W_a^{\parallel}$ when fitting SM.

**Limitations and future extensions**

Our study focused exclusively on the effects of microstructure with scalar susceptibility on the MRI signal phase. However, we did not account for susceptibility anisotropy of myelin or other potential field perturbing sources such as iron-filled neuroglia[75], which may affect the mesoscopic contribution to the MRI signal phase. For example, previous simulations[76] of the field from orientationally dispersed cylinders with susceptibility anisotropy, even for perfect cylinders discretized on a 3D grid with high resolution led to an error of around 5% compared to analytical results. We have previously shown how spheres may also play an important role in the measured frequency shift[76], as their positions can correlate with the axons both inside and outside the axons. However, if the sources have sufficiently high magnetic susceptibility, a different model picture accounting for strong static dephasing[8] may be necessary. We therefore plan to add additional complexity to the simulation in the future, e.g., by utilizing other publicly available EM data with segmented neuroglia, vessels and other relevant structures.[77]

We only considered the intra-axonal signal here for several reasons. First, preparing the tissue for EM may alter the geometry of the extra-axonal space. Also, the extra-axonal space has a complex boundary, which makes appropriate boundary conditions difficult to implement. Only a fraction of the intra-axonal volume was used for MC simulation, which was mainly a result of rejecting axons shorter than 20 μm. While this might give the impression of a lower g-ratio $g \approx \sqrt{1/(1 + \zeta_m/\zeta_a)}$, using the raw intra-axonal and myelin volume fraction lead to a mean g-ratio of around 0.59 for SHAM and 0.54 for TBI, which agrees with previous studies[67]. This also shows that our results are valid for a range of myelin thicknesses as expected[12,13] as we perform simulations in substrates with various thicknesses.

*Diffusion-weighting*

Here we simulated diffusion weighting with ideal pulsed diffusion gradients and a perfect 180-degree RF pulse. While this differs from a real measurement, our simulations provide important first insights into the dMRI signal from such a large mesoscopic in-silico sample. Future work will extend the simulation to include realistic diffusion gradient shapes, but also B-tensor encoding like double diffusion encoding[78]. In our simulations, we made the standard approximation of neglecting the permeability of myelin, which may change the diffusion effects. However, we suspect this to have a minor impact on the estimation of the fODF. Recent studies have also demonstrated the potential of using diffusion weighting to filter different axonal bundles according to their orientation to the external magnetic field.[76] Given the natural orientation dispersion of axons, this approach may enable estimating orientation dependent transverse relaxation across echo times $T_E$, and the signal dephasing and transverse relaxation as functions of the delay time $\Delta T_E$ of the asymmetric spin echo sampled at times after $T_E$. While this has so far been validated for ideal cylinder phantoms[76], it has not been investigated in realistic substrates. We therefore intend to investigate this approach using this MC simulation framework in a future study.

*Computational demand*

The random walk simulation was carried out in double precision as high diffusion times led to catastrophic cancellation in single precision. However, the internal field tensor $\boldsymbol{\varphi}$ could be stored in single precision, which allocated around 32 Gb RAM. For the MC simulation, this meant phase accumulation due to internal fields had to be carried out on the CPU rather than GPU, as our desktop GPU only had 12Gb of memory. A more powerful GPU could therefore speed up the computation time substantially. Allocating the full frequency tensor $\mathbf{A}(\boldsymbol{r})$ allows for fast extraction of values, as particle positions can be converted into an index position in the array. This basically makes finding the added phase for every particle a single operation. As the frequency shift is only considered inside axons, the low axonal volume fraction of approximately 15% made the problem somewhat sparse. However, only storing non-zero values require storing their index. This means that each particle position must be matched with the correct index in the array, resulting in two operations (a search and an extraction), ultimately leading to higher computation time in comparison to storing the full array, and directly extracting the values. However, better methods might be possible (e.g., utilizing the Cusparse library in CUDA), but this was not pursued here.

**Propositions for QSM and STI**

Here we showed that estimating the mean magnetic field from the gradient echo signal $S_{\text{MGE}}(t)$ must be done with consideration of the magnitude of the frequency shift described by the tissue magnetic susceptibility, magnetic field strength and readout times. The best estimation achieved in our study for 7T was for times less than 30 ms; at 3T, on the other hand, we could successfully estimate the mean magnetic field for all times considered. However, since water close to the edge of the axons, including myelin water, likely affects the Larmor frequency shift at low echo times[79–82], it is favorable to measure $S_{\text{MGE}}(t)$ at times longer than the transverse relaxation time of myelin water. This is necessary as long as a complete model of myelin water remains unsolved[83]. Since measuring $S_{\text{MGE}}(t)$ for long echo times (cf. Figure 5) might hamper estimating the mean magnetic field directly, we found that a better strategy may be to instead perform a spin echo experiment with high $T_E$, and then sample its decay, e.g., by sampling an asymmetric spin echo at times $\Delta T_E$. As shown in Figure 6, this greatly improved estimation of the mean magnetic field. However, using this approach cannot be carried out at arbitrarily high $T_E$. The longer $T_E$, the greater the signal attenuation due to transverse relaxation, which ultimately changes the compartmental weighting of the frequency distribution. In our simulations, we found $T_E$ up to 80 ms could estimate the mean magnetic field at 7T reasonably, by sampling the SE signal after $T_E$ at times $\Delta T_E$ up to 20 ms. The maximum timings used in both MGE and ASE signals depends ultimately on the variance in the frequency shifts across the different axons. This is why SHAM-49-ipsi produced the best estimates, as here only cingulum was included. We therefore propose that susceptibility estimation may benefit from using the signal phase of an asymmetric spin echo to avoid contributions from myelin water. In addition, a dMRI dataset with high b-values (around b = 5 ms/um$^2$ or higher in-vivo) and multiple gradient directions (around 60 directions) should be acquired with a diffusion time Δ around 40 ms, which can be achieved in around 10-20 minutes on a 3T clinical scanner. A potential strategy to only measure the MRI phase of intra-axonal water, which was mainly investigated here, may be to include delayed read-outs[14,84–87] of the dMRI dataset with high b-value, average the complex data across gradient directions (a powder average) in order to remove the signal contribution from extra-axonal water[88].

# 6| Conclusion

We performed Monte-Carlo simulations in mesoscopically sized samples of white matter axons to test if structural information obtained with diffusion MRI can be used to account for the effect of microstructure probed with the susceptibility-induced Larmor frequency shift of the gradient echo signal. We found that by fitting the Standard Model of diffusion in white matter to the simulated pulsed gradient spin-echo signal, the estimated fiber orientation distribution function could successfully be used to account for the orientation dependence of the Larmor frequency shift with respect to the B0 field. Our simulations also showed that intra-axonal axial kurtosis affects the diffusion-weighted MRI signal and should not be neglected.

Our work is an important step towards validating the use of dMRI information to account for sub-voxel frequency shift caused by structural anisotropy of WM, which is crucial for robust susceptibility estimation – with the end goal of developing a unified model of both susceptibility and diffusion effects in MRI.

# 7| Acknowledgements

This study is funded by the Independent Research Fund Denmark (Grant number 10.46540/3103-00144B). The data that support the findings of this study are available from the corresponding author upon reasonable request.

# 8| Supporting Material

**S1) Total Larmor frequency shift and Extra-axonal Larmor frequency shift of SHAM-ipsi-25**

Here we show for SHAM-ipsi-25 that the first signal cumulant, describing the mean magnetic field, is also measurable with MRI for extra-axonal water $\overline{\Omega}_e$, and when considering both intra-axonal and extra-axonal water denoted $\overline{\Omega}$. For extra-axonal water, we performed Monte-Carlo simulations similar to what is described in the main text for intra-axonal water, but here we used a reflective boundary condition when extra-axonal particles reached the edges of the volume. Second, the myelin sheath was synthetically generated by dilating the intra-axonal mask and then subtracting the original intra-axonal mask. The dilation magnitude was chosen such that the g-ratio was close to 0.6. Figure 11 shows the mean ratio $\langle \overline{\Omega}_{MGE}(\mathbf{B}_0)/\overline{\Omega}_e(\mathbf{B}_0)\rangle_{\hat{\mathbf{B}}}$ and standard deviation (SD) for extra-axonal water only, or $\langle \overline{\Omega}_{MGE}(\mathbf{B}_0)/\overline{\Omega}(\mathbf{B}_0)\rangle_{\hat{\mathbf{B}}}$ for both intra- and extra-axonal water, between the linear term $\overline{\Omega}_{MGE}(\mathbf{B}_0)$ found from fitting the MC-simulated MGE signal phase to either a linear or 3rd order polynomial, and $\overline{\Omega}_e(\mathbf{B}_0)$ or $\overline{\Omega}(\mathbf{B}_0)$, respectively. Both the mean and standard deviation are calculated over the

$\hat{\mathbf{B}}$ directions, and the ratio is shown for different maximum times $t_{\max}$ and $(\Delta T_E)_{\max}$ used in the fit for MGE and ASE, respectively. For 3T, we see that the first cumulant is well estimated in all cases, no matter if we use linear or polynomial fitting. At 7T, however, the standard deviation increases. At 7T, using the ASE phase and polynomial fitting produced a good estimation of the first cumulant. Figure 12 shows that our theoretical prediction agrees with the mean magnetic field across all water.

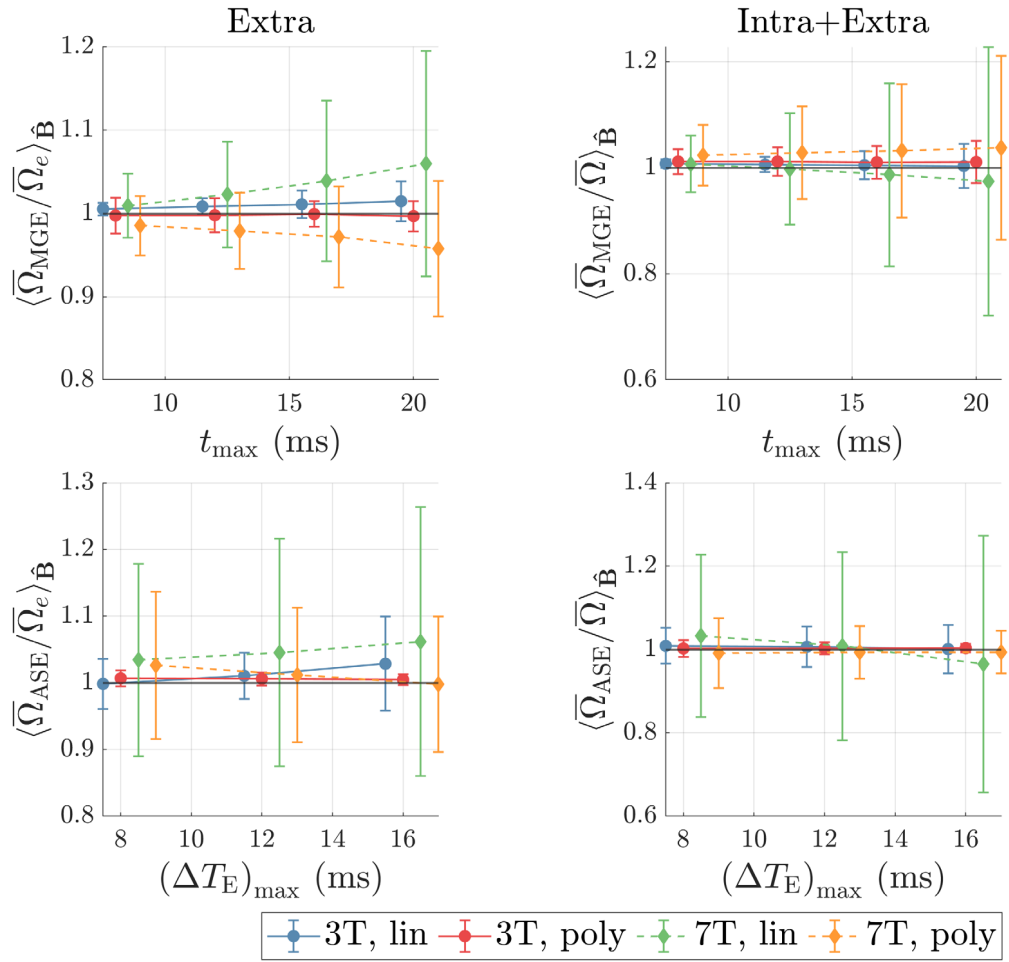

*Figure 11 - Error ratio from fitting a linear or 3rd order polynomial to the phase of the MC simulated MGE and ASE signal for SHAM-ipsi-25. Colors indicate different fitting protocols and field strengths. The x-axis indicates the maximum time used to fit the phase*

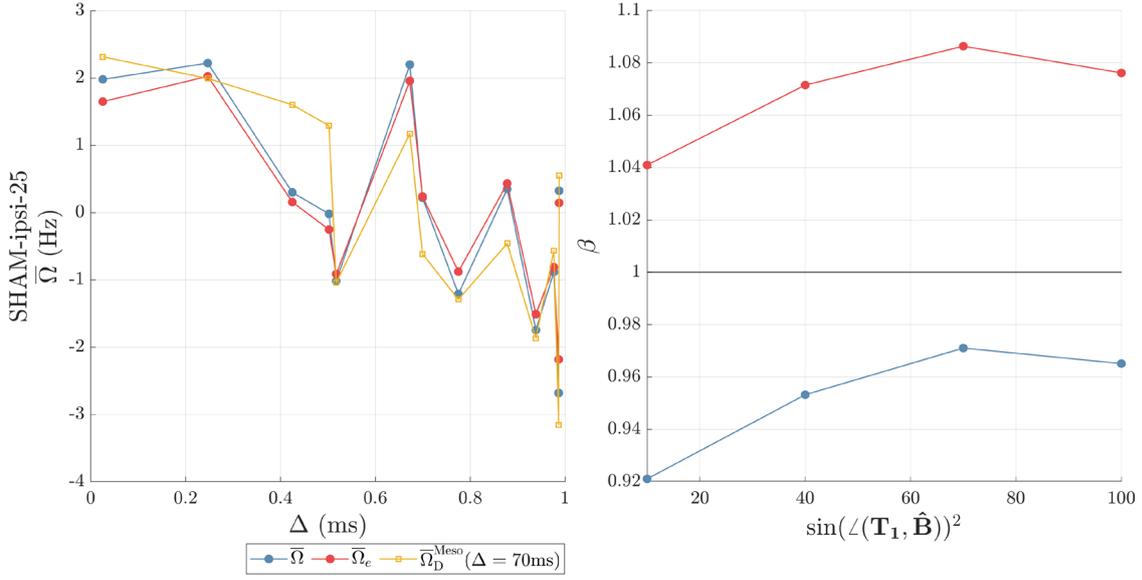

*Figure 12 – **Right**: Normalized susceptibility fitting parameter β for all WM substrates. β was estimated by minimizing the squared difference between the mean Larmor frequency shift corresponding to the mean magnetic field and the estimated mesoscopic frequency $\overline{\Omega}^{Meso}$. The frequency was estimated using the fODF from fitting the standard model of diffusion with intra-axonal axial kurtosis $W_a^{\parallel}$ to the simulated PGSE signal. The x-axis shows the time Δ controlling the width of the coarse graining filter, and corresponds to the time between the gradient pulses for the PGSE signal. **Left**: Mean Larmor frequency shift. Blue curve shows the total Larmor frequency shift described by the mean magnetic field, while the red shows the intra-axonal frequency shift $\overline{\Omega}_a$, while the yellow shows the estimated Larmor frequency shift $\overline{\Omega}_D^{Meso}$ based on the model for dispersed cylinders and estimated using the fODF estimated from MC simulated diffusion MRI signal with diffusion time Δ = 70 ms. The x-axis shows $\sin^2$ to the angle between the external field $\hat{\mathbf{B}}$ and the principal eigenvector $\mathbf{T_1}$ of the scatter matrix $\mathbf{T}$.*

## S2) Validation of Monte-Carlo simulation

Here we present a set of Monte-Carlo (MC) simulations to ensure that the MC simulations in the main manuscript are trustworthy.

a) Single cylinder

We considered the time dependent axial and radial diffusivity $D_\parallel(t), D_\perp(t)$ and kurtosis $K_\parallel(t), K_\perp(t)$

$$D(\hat{\mathbf{n}}; t) = \frac{\langle (\mathbf{r} \cdot \hat{\mathbf{n}})^2 \rangle}{2t} \quad (20)$$

$$K(\hat{\mathbf{n}}; t) = \frac{\langle (\mathbf{r} \cdot \hat{\mathbf{n}})^4 \rangle}{\langle (\mathbf{r} \cdot \hat{\mathbf{n}})^2 \rangle^2} - 3 \quad (21)$$

at diffusion time *t* of a perfect cylinder and compared it with known analytical solutions. The subscripts ∥, ⊥ correspond to using $\hat{\mathbf{n}}$ parallel or perpendicular to the cylinder axis, respectively. The purpose of the simulation is also to check that the axial diffusivity $D_\parallel$ is biased from using rejection sampling, i.e., if rejecting the step when a particle collides with the cylinder boundary reduces the diffusivity depending on the surface-to-volume ratio S/V and diffusion step length $\delta l$. We discretized a single cylinder with a radius of 1 μm and length 68 μm (comparable S/V to the real axons) inside a 3D grid with a resolution of 0.1 μm³ – similar to the large-scale simulations. Around $5 \cdot 10^3$ particles inside the cylinder was used to have the same density of particles was as in the axons. Figure 11 shows the axial and radial diffusivities $D_\parallel(t), D_\perp(t)$ and kurtoses $K_\parallel(t), K_\perp(t)$ from both MC and analytical solutions. We also plotted the analytical solutions for $D_\perp(t)$ and $K_\perp(t)$ for comparison with MC, while for $D_\parallel(t)$, we also plot the time-averaged axial diffusivity $\langle D_\parallel \rangle_t$ and the axial diffusivity, expected to be biased according to

$$D_0' \approx D_0 \left(1 - \frac{3}{16}\frac{S}{V}\delta l\right) \tag{22}$$

due to using rejection sampling, where $\delta l S/V = 0.94$. We see in Figure 11 that $\langle D_\parallel \rangle_t \approx 0.94 D_0 \approx D_0'$ was biased at around 6% in agreement with the expectation. The axial kurtosis quickly converged to zero as expected, while the radial diffusivity and kurtosis followed the analytical solution. The axial variance is around 2%, which is indicative of the low number of particles inside the cylinder. However, increasing the number of axons lowers the variance. Hence, if a single axon was to be considered, the number of particles should be increased, but pooling the results over many, we found this particle density to be sufficient, as will be evident in the next section.

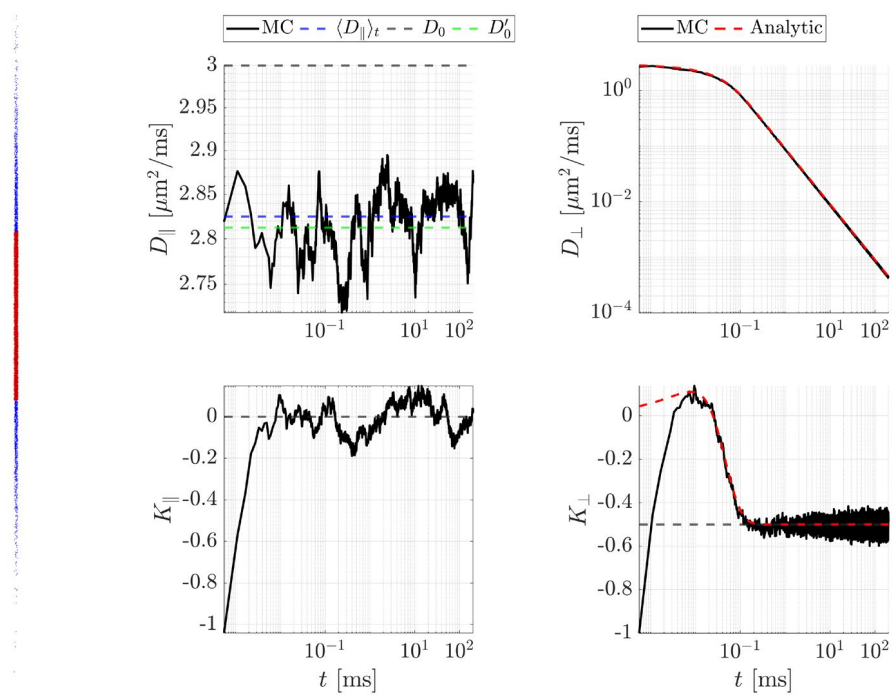

Figure 13 – Monte-Carlo simulation of intra-cylindrical diffusion: To the left, red dots indicate particles positions inside the cylinder, where a particle passing through one end comes out the other end. Blue dots represent the actual displacement at $t = 200$ ms. First column shows the axial diffusivity $D_\parallel(t)$ and kurtosis $K_\parallel(t)$, while the second column shows the radial diffusivity and kurtosis $D_\perp(t), K_\perp(t)$, respectively, including analytical solutions in red.

b) Realistic axons

Here we simulated $D_\parallel(t), D_\perp(t), K_\parallel(t), K_\perp(t)$ in realistic axons to investigate their asymptotic behaviors for $t \to \infty$. In particular, the power law scaling should be $D_\parallel(t), K_\parallel(t) \sim 1/\sqrt{t}$ while $D_\perp(t), K_\perp(t) \sim 1/t$ for $t \to \infty$.[89–91] We compute the time dependence for the diffusion and kurtosis parameters for 945 different axons from the cingulum bundle in their own frame of reference $\hat{\boldsymbol{n}}_i$ for the i'th axon. The axial direction $\hat{\boldsymbol{n}}_i$ was chosen by averaging the tangent directions of the center-of-mass line of an axon, coarse-grained by a Gaussian filter corresponding to a diffusion length of 25 μm. We computed the total diffusivity and kurtosis $D(t), K(t)$ in the common frame of reference defined as $\hat{\boldsymbol{n}} = \sum_i f_i \hat{\boldsymbol{n}}_i$ to include the natural orientation dispersion

$$D(t) = \sum_i f_i D_i(\hat{\boldsymbol{n}}_i; t) \qquad (23)$$

$$K(t) = \frac{1}{D(t)^2} \sum_i f_i \big(D_i(\hat{\boldsymbol{n}}_i; t) - D(t)\big)^2 + f_i D_i(\hat{\boldsymbol{n}}_i; t)^2 K_i(\hat{\boldsymbol{n}}_i; t) \qquad (24)$$

We considered here diffusion times up to $t = 400$ ms. We also calculated with a density of particles of either 72 particles/μm³ or 572 particles/μm³, where the first corresponds to the density used in the main simulations. This is to demonstrate that we do not need a lot of particles inside every axon to produce robust results. This is because we are not interested in the diffusivity of individual axons but rather the behavior over the entire population. Figure 12 shows the decay of $D_\parallel(t), D_\perp(t)$ and kurtoses $K_\parallel(t), K_\perp(t)$ for long diffusion times for individual axons and all of them. Figure 12A is for axons in their own reference frame, while Figure 12B is in their common reference frame. Our results agree with previous studies[89–91] when every axon is considered in their own reference frame. Surprisingly, when we compute $K_\perp(t)$ in the common frame of reference, it diverges even though $K_i(\hat{\boldsymbol{n}}; t)$ does not. Hence the divergence comes from the diffusion variance from certain axons $D_i(\hat{\boldsymbol{n}}_i; t) - D(t)$ converge to 0 at a slower rate as $D(t)$. Figure 13 shows the contribution to $K_\perp(t)$ from $D_i(\hat{\boldsymbol{n}}_i; t)$ when calculated in the common reference frame, and in the individual. Axons that deviate from the main orientation can clearly contribute to the diffusion variance such that $K_\perp(t)$ appears to diverge within the considered time scale. Looking at the total parallel kurtosis $K_\parallel(t)$ in the individual axon reference frame, we observed that it remains non-zero even for very high diffusion times. For that reason, we also considered including $W_\parallel$ when fitting the Standard Model of diffusion to the MC simulated PGSE signal.

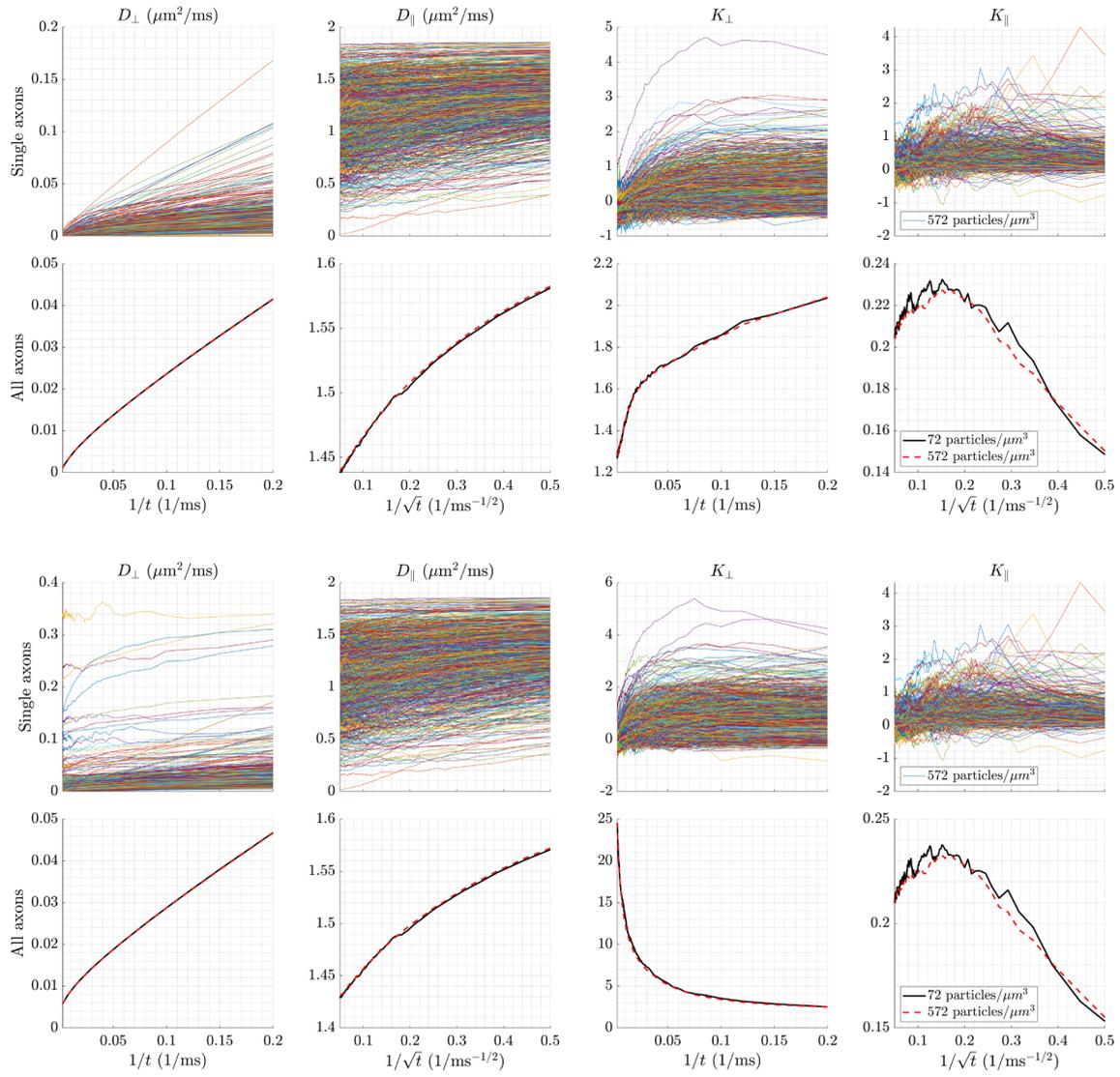

Figure 14 – Monte-Carlo simulation of intra-axonal diffusion and kurtosis: The axial diffusivity $D_\parallel(t)$ and kurtosis $K_\parallel(t)$ are plotted as a function of $1/t$ while $D_\perp(t), K_\perp(t)$ are plotted as a function of $1/\sqrt{t}$. The first and third row show single axons, while the second and fourth row show all. The two first rows show when the diffusivity and kurtosis are calculated with axons in their own reference frames, while the third and fourth row show for axons calculated in their common reference frame.

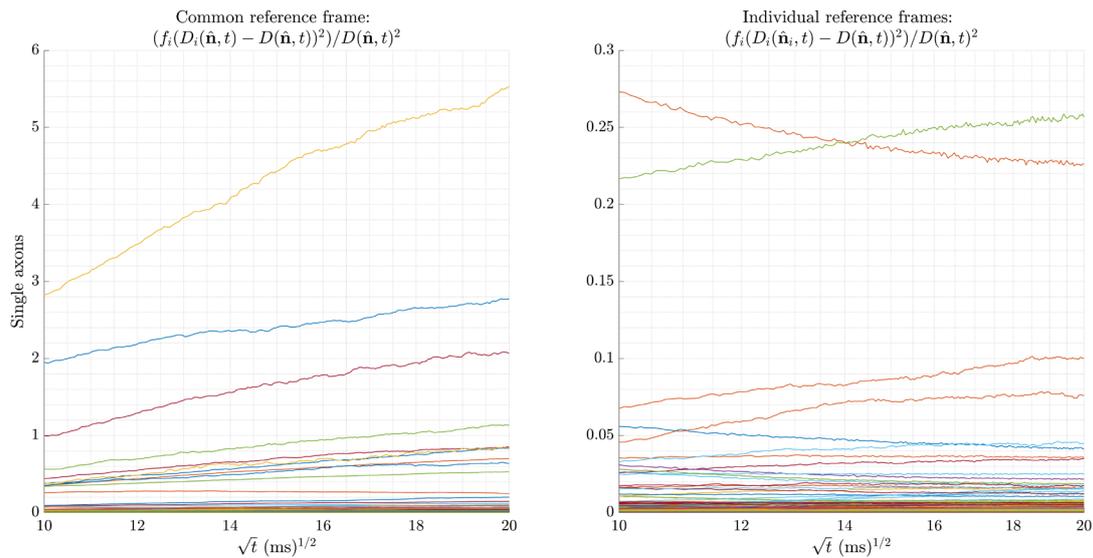

*Figure 15 – Contribution to intra-axonal diffusion variance from each axon: The first plot shows the individual contributions to the diffusion variance contribution to the kurtosis from all 945 axons calculated in their common reference frame. Axons deviating from the common reference frame decays slower than the total mean diffusivity $D(\hat{\mathbf{n}}, t)$, which leads to a diverging kurtosis contribution. The second plot shows each contribution to the diffusion variance in each axon's own reference frame. Here, most axon tend towards zero.*

# 9| References